\documentclass[%
prb,
amsmath,amssymb,
twocolumn,
reprint,
superscriptaddress,
floatfix
]{revtex4-2}

\usepackage{graphicx} 
\usepackage{bm}
\usepackage{color}
\usepackage{epstopdf}
\usepackage{enumitem}
\usepackage[utf8]{inputenc} 
\usepackage{textcomp} 
\usepackage[normalem]{ulem}

\usepackage[FIGTOPCAP,hang, nooneline]{subfigure} 

\usepackage[colorlinks=true,allcolors=blue]{hyperref}
\usepackage{orcidlink} 

\definecolor{Micolor1}{RGB}{0,128,0}

\begin{document}

\title{Chiral molecule-induced contributions to ferromagnetic resonance}

\author{J\"urgen Lindner\,\orcidlink{0000-0002-4955-515X}}
\affiliation{Helmholtz-Zentrum Dresden--Rossendorf, Institute of Ion Beam Physics and Materials Research, Bautzner Landstra\ss e 400, 01328 Dresden, Germany}

\author{Pedro Contreras-Gallardo\,\orcidlink{...}}
\affiliation{Universidad Técnica Federico Santa María, Departamento de Física, Avenida España 1680, Valparaíso, Chile} 

\author{Abhishek Singh} 
\affiliation{Helmholtz-Zentrum Dresden--Rossendorf, Institute of Ion Beam Physics and Materials Research, Bautzner Landstra\ss e 400, 01328 Dresden, Germany} 

\author{Ruslan Salikhov\,\orcidlink{0000-0001-8461-0743}} 
\affiliation{Helmholtz-Zentrum Dresden--Rossendorf, Institute of Ion Beam Physics and Materials Research, Bautzner Landstra\ss e 400, 01328 Dresden, Germany} 

\author{Anna Semisalova\,\orcidlink{....}} 
\affiliation{Universit\"at Duisburg--Essen, Fakult\"at f\"ur Physik, Lotharstra\ss e 1-21, 47048 Duisburg, Germany} 

\author{Olav Hellwig\,\orcidlink{0000-0002-1351-5623}} 
\affiliation{Helmholtz-Zentrum Dresden--Rossendorf, Institute of Ion Beam Physics and Materials Research, Bautzner Landstra\ss e 400, 01328 Dresden, Germany} 
\affiliation{Institute of Physics, Chemnitz University of Technology, 09107 Chemnitz, Germany}

\author{Rodolfo Gallardo\,\orcidlink{0000-0002-4495-5592}}
\affiliation{Universidad Técnica Federico Santa María, Departamento de Física, Avenida España 1680, Valparaíso, Chile} 

\author{Anna Lewandowska-Andralojc\,\orcidlink{0000-0003-3166-2640}} 
\affiliation{Faculty of Chemistry, Adam Mickiewicz University, Uniwersytetu Poznanskiego 8, 61-614 Poznan, Poland} 

\author{Kilian Lenz\,\orcidlink{0000-0001-5528-5080}} 
\affiliation{Helmholtz-Zentrum Dresden--Rossendorf, Institute of Ion Beam Physics and Materials Research, Bautzner Landstra\ss e 400, 01328 Dresden, Germany} 

\author{Aleksandra Lindner\,\orcidlink{0000-0002-5119-7769}}
\affiliation{Helmholtz-Zentrum Dresden--Rossendorf, Institute of Ion Beam Physics and Materials Research, Bautzner Landstra\ss e 400, 01328 Dresden, Germany}

\newcommand{\p}{\partial}
\newcommand{\directions}[1]{{$\langle#1\rangle$}}
\newcommand{\ipdirections}[1]{{$\langle#1\rangle_\mathrm{ip}$}}
\newcommand{\direction}[1]{{[#1]}}
\newcommand{\ipdirection}[1]{{[#1]$_\mathrm{ip}$}}
\newcommand{\plane}[1]{{(#1)}}
\newcommand{\Ms}{M_\mathrm{s}}
\newcommand{\Meff}{\mu_0M_\mathrm{eff}}	
\newcommand{\KuparM}{2K_{2\parallel}/M_\mathrm{s}}
\newcommand{\KuperpM}{2K_{2\perp}/M_\mathrm{s}}
\newcommand{\KcparM}{2K_{4\parallel}/M_\mathrm{s}}
\newcommand{\Kupar}{K_{2\parallel}}
\newcommand{\Kuperp}{K_{2\perp}}
\newcommand{\Kcpar}{K_{4\parallel}}
\newcommand{\Kvol}{K^\mathrm{vol}}
\newcommand{\Kint}{K^\mathrm{int}}
\newcommand{\Bres}{B_\mathrm{res}}

\begin{abstract}

Despite extensive research on chirality-driven spin selectivity, most studies have focused on static magnetic properties, while the influence of chirality on the dynamic magnetic response remains largely unexplored.
Here, we investigate how chiral molecular interfaces affect magnetization dynamics in thin Co/Ni multilayers with perpendicular magnetic anisotropy using broadband ferromagnetic resonance spectroscopy. A comparison between bare (reference) films and molecule-functionalized (hybrid) samples reveals no measurable changes in either the resonance field or the linewidth that could be attributed to the presence of the chiral environment.
Motivated by our findings we develop a macrospin description that distinguishes equilibrium modifications of the magnetic free-energy landscape (MIPAC-type effects) from non-equilibrium, CISS-induced spin torques. Our analysis shows that equilibrium modifications primarily shift the resonance condition via changes to the free energy landscape and thereby the effective field, whereas damping-like non-equilibrium torques provide a distinct channel for varying the effective damping rate. This approach establishes clear criteria for disentangling chiral-interface-induced energy modifications from torque-driven dynamical effects in ferromagnetic resonance experiments. 

\end{abstract}

\maketitle

\section{Introduction} \label{sec_1}

For a long time, spin-selective electron transfer and transport were considered phenomena restricted to either inorganic magnetic materials or inorganic systems with exceptionally strong spin–orbit coupling \cite{sinova_spin_2015}.

This view changed with the publication of a Science paper \cite{ray_asymmetric_1999} reporting the generation of spin-polarized electron beams when photoelectrons, emitted from thin gold films, passed through a thin layer of chiral molecules adsorbed on the gold surface. 
A few years later, spin polarizations exceeding 60\% were reported for electrons transmitted through self-assembled monolayers of double-stranded DNA \cite{gohler_spin_2011}.

The preferential transmission of electrons with either spin-up or spin-down orientation through chiral systems, depending on the handedness of the molecules, direction of electron flow and the orientation of the molecular electric dipole moments relative to the surface, became known as the ``chiral-induced spin selectivity'' (CISS) effect \cite{naaman_chiral-induced_2012, naaman_chiral_2020}.

At the same time, closely related phenomena have been observed upon adsorption of chiral molecules on the ferromagnetic surfaces, where the magnetic properties of the metal are modified---in some cases even leading to magnetization switching. This effect was termed ``magnetism-induced by the proximity adsorption of chiral molecules'' (MIPAC) \cite{koplovitz_single_2019, ben_dor_magnetization_2017}.
It arises from interfacial charge transfer processes accompanying adsorption of the chiral molecules on magnetic surfaces. As such, it is often not clearly distinguished from the observed CISS effect itself.
Meanwhile, chirality-driven spin selectivity has been observed in a wide range of materials, spanning oligopeptides and DNA, simple amino acids, metal–organic frameworks, and chiral perovskites \cite{bloom_chiral_2024}.
It has also been shown not to be restricted to molecular layers in direct contact with a metallic surface, but to occur in organic molecules in which donor and acceptor moieties are connected by a chiral linker \cite{eckvahl_direct_2023}.
It has been used to separate enantiomers on achiral perpendicularly magnetized magnetic substrates, where enantioselective recognition was achieved through exchange interactions \cite{banerjee-ghosh_separation_2018}.
Moreover, it has enabled the helical aggregation of achiral components without the use of a chiral catalyst \cite{aizawa_enantioselectivity_2023}.

Although the first studies of spin selectivity related to molecular chirality were carried out using photoelectron spectroscopy \cite{gohler_spin_2011, mollers_chirality_2022}, by now numerous other techniques have been employed to detect spin polarization induced by chiral materials \cite{bloom_chiral_2024, naaman_spintronics_2015}.
Among the most widely used are transport measurements, which require a charge flow during the experiment \cite{bloom_chiral_2024}.
In such cases, magneto-resistance can be measured either for ensembles of molecules (e.g., Hall bars or spin valves) or at the single-molecule level using magnetic conductive atomic force microscopy or scanning tunneling microscopy \cite{aragones_magnetoresistive_2022, albro_measurement_2025, hossain_observation_2025, nguyen_mechanism_2024, nguyen_cooperative_2022, eckshtain-levi_cold_2016}.
Spin-dependent electrochemical methods likewise require electron transfer between the chiral molecules and the electrode interface \cite{einati_light-controlled_2015, mondal_field_2015}.
In the absence of electron flow, the influence of chirality on an adjacent ferromagnet can be probed using magnetic force microscopy, magnetometry, and magneto-optical techniques \cite{ben_dor_magnetization_2017, bloom_chiral_2024, sharma_control_2020}.

Despite very intensive research on chirality-driven spin selectivity, significant discrepancies remain between experimental results and their theoretical interpretations. 
There is still no generally accepted mechanism for either the CISS or the MIPAC effect \cite{bloom_chiral_2024}.

For instance, it has been shown that asymmetric two-terminal magneto-resistance in the nonlinear transport regime is a more general phenomenon and can also be observed when an achiral medium is contacted by nonmagnetic electrodes \cite{hossain_observation_2025}.
Likewise, large magneto-resistance values, often widely interpreted as evidence of spin-dependent transport, may alternatively be explained by the experimentally demonstrated dependence of the electrostatic contact potential between the chiral materials and a ferromagnet on both the magnetization direction and molecular chirality. 
In this scenario, the transport barrier is already strongly modified under equilibrium conditions, i.e., in the absence of any bias current \cite{tirion_mechanism_2024}. 
Investigations of the CISS effect in chiral paramagnetic radicals adsorbed on gold-capped Ni surfaces, using spin-dependent electrochemistry and charge transport measurements, reveal that in these systems the dominant contribution to the observed spin polarization arises from radical spin polarization rather than from molecular chirality \cite{sousa_interrogating_2024}.

Ferromagnetic resonance spectroscopy (FMR) is widely used to investigate spin-related processes at interfaces \cite{lindner_applications_2012}.
By directly probing the magnetic response of a system under non-equilibrium conditions, it offers greater sensitivity than methods that rely solely on static magnetic properties. 
It is therefore striking that most studies on CISS and MIPAC systems--despite their inherently interfacial nature--have so far focused on static magnetic properties of ferromagnets in contact with chiral molecules, rather than on the magnetization dynamics of these systems. The available reports \cite{hatajiri_gilbert_2024, sun_inverse_2024, moharana_chiral-induced_2025, sun_colossal_2024} discussing magnetization dynamics in chiral/ferromagnetic hybrids are shortly described below.

Time-resolved magneto-optic Kerr effect (TR-MOKE) was used to probe the dynamic magnetic properties of 5-nm-thick permalloy films deposited on approximately 60-nm-thick 2D chiral perovskite films in a charge-current-free environment \cite{hatajiri_gilbert_2024}. 
A laser was used to excite the magnetization dynamics in permalloy, and subsequent spin-pumping into the organic layer was observed. 
The reports revealed only a minor variation of Gilbert damping constant depending on the magnetization direction of the permalloy. 
The absence of significant chirality-induced damping is surprising, given the large spin polarizations reported in the literature. 

In another study \cite{sun_inverse_2024}, a 15-nm-permalloy layer was deposited on a thin film of a chiral $\pi$-conjugated polymer, and spin-pumping measurements were performed. 
The magnetization dynamics in the permalloy were excited by a microwave field, and the resulting spin-current injected into the polymer layer was detected through inverse spin-Hall voltage measurements. 
In contrast to the previously described TR-MOKE measurements, a strongly enhanced damping was observed. 
This effect was interpreted as evidence of a chirality-induced unconventional spin–orbit coupling arising from the helical structure of the polymers. 

A similar measurement principle was used to investigate the impact of chirality on pure spin-currents generated in yttrium iron garnet and injected into helical polyalanine molecules adsorbed on a 4 nm gold capping layer on the YIG surface \cite{moharana_chiral-induced_2025}.
The direction of the spin-current's polarization was controlled by an external magnetic field. 
A significant change in the magnitude of the inverse spin-Hall voltage was observed upon reversal of the magnetic field polarity, whereas no such effect was detected in reference samples without chiral molecules. 
Moreover, the sign of this change was reversed for molecules of opposite chirality, and the maximum (minimum) spin selectivity of about 50--60\% was achieved, when the spin polarization was parallel (antiparallel) to the chiral axis \cite{moharana_chiral-induced_2025}.

Likewise, spin-pumping was applied to investigate whether chiral cobalt oxide can act as a spin analyzer \cite{sun_colossal_2024}. For this purpose, a 15-nm-permalloy film was excited by microwaves, and the subsequent absorption of the generated spin-current by the adjacent chiral layer was detected through voltage measurements in a Hall device. 
The authors reported an anomalous spin-current absorption by the cobalt oxide, evidenced by colossal anisotropic damping with twofold symmetry. The differences in damping anisotropy between the two opposite handednesses of cobalt oxide were negligible, whereas a substantial difference was observed between the chiral and achiral samples \cite{sun_colossal_2024}.

\begin{figure}[tb] 
    \centering
    \includegraphics[width=0.8\columnwidth]{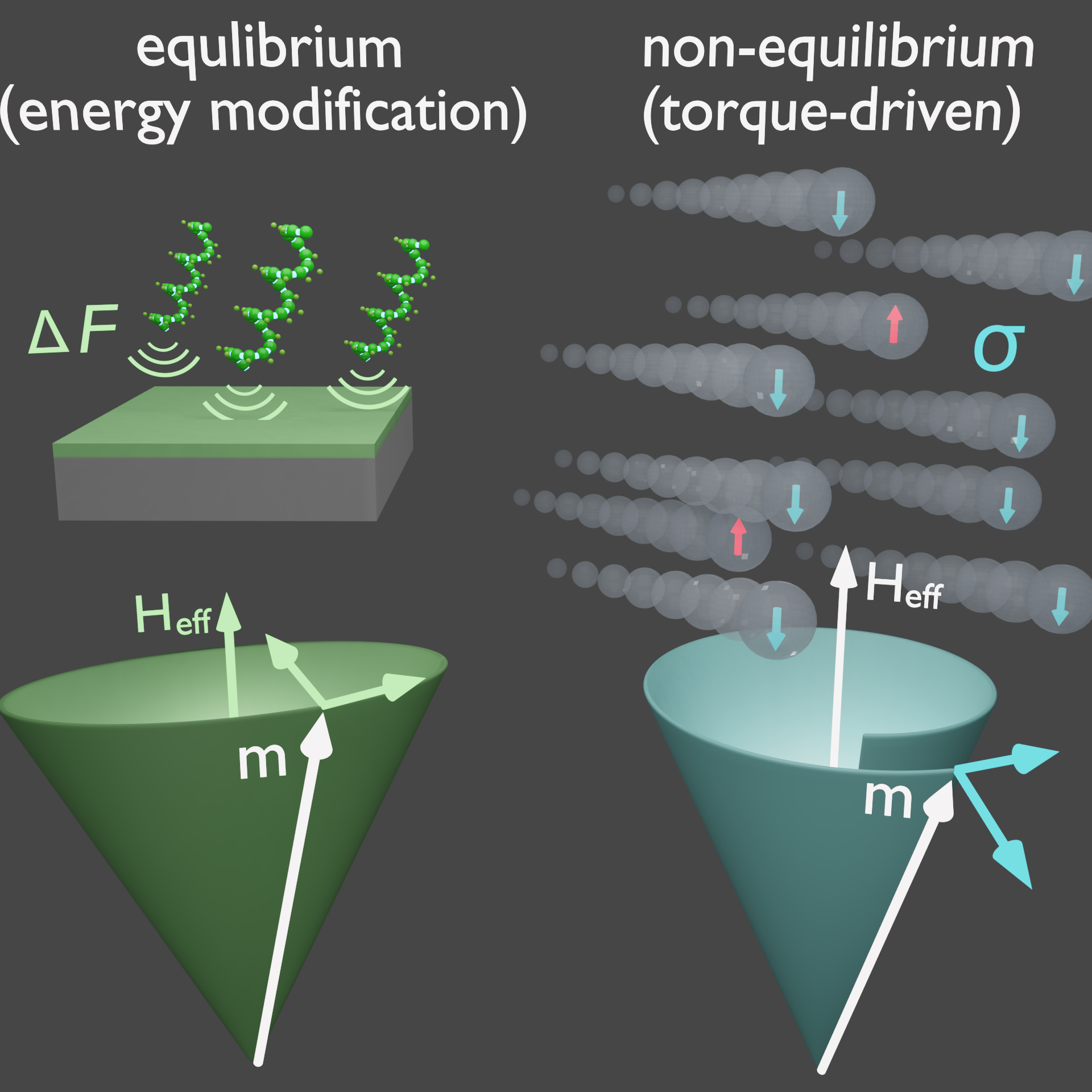}  
    \caption{Schematic representation of equilibrium and non-equilibrium chiral-interface effects on spin dynamics. In equilibrium (left), chiral molecules modify the magnetic free-energy density, $\Delta F(\bm M)$, thereby altering the effective field $\bm H_{\mathrm{eff}}$ and the underlying energy landscape. Under non-equilibrium conditions (right), a spin polarization $\hat{\bm \sigma}$ generates field-like and damping-like torques. The latter enables energy transfer into or out of the magnetic system and modifies the precession cone angle. Green and cyan colors highlight the quantities primarily affected in the two cases.}
    \label{fig:principle} 
\end{figure}

Interestingly, in all of the above-mentioned studies reporting a strong influence of chiral molecules on spin-currents generated in the FM layer, electrical detection was employed. That is, changes in the dynamic magnetic properties of the FM were always accompanied by electron flow in the system,indicating the involvement of magneto-electric effects. 
Moreover, the TR-MOKE results, in which negligible chirality-induced damping was observed, contradict the findings of the electrical measurements, which showed the opposite behavior.

Therefore, further studies of magnetization dynamics in chiral/ferromagnetic hybrids are required that \emph{directly} probe the magnetic properties of the FM in the vicinity of the chiral organic layer, rather than relying on electrical detection. 
Moreover, disentangling the dynamic magnetic behavior of the FM in a chiral environment from any electron flow in the system is essential for obtaining detailed insight into the influence of chirality on spin dynamics in such hybrids and for properly addressing several open questions related to CISS and MIPAC. 

Among the most important of these is whether chirality can indeed modify the magnetic properties of the FM, and to what extent electron transport is necessary for such effects to occur. 
Addressing these issues requires not only carefully designed experiments, but also theoretical approaches capable of predicting and explaining how CISS and MIPAC effects are expected to influence magnetization dynamics in the first place.
In the study presented here, we address this genuine gap by performing systematic broadband FMR measurements on thin ferromagnetic films with perpendicular magnetic anisotropy, capped with a gold layer that serves as an anchoring platform for chiral organic molecules.

In the present work, we first discuss FMR results on Co/Ni multilayer thin films with perpendicular magnetic anisotropy (PMA), onto which oligopeptides modified with a porphyrin chromophore were covalently attached via gold–sulfur bonds. Secondly, as FMR is a method rarely used so far to detect CISS or MIPAC effects, we supplement our experimental findings with a detailed theoretical analysis of the impact of chiral-interface-induced contributions to ferromagnetic resonance. We consider both, non-equilibrium spin-polarization effects, described by the unit vector $\hat{\bm \sigma}$ generating field-like and damping-like torques, and equilibrium modifications of the magnetic free-energy density, as schematically depicted in Fig. \ref{fig:principle}. Both contributions can be treated within a common framework, in which the geometric structure of the torque and the curvature of the free energy determine how these effects manifest in the FMR response. This provides a general basis to distinguish torque-driven non-equilibrium effects from equilibrium energy modifications and to identify CISS- and MIPAC-type signatures in FMR experiments. 

\section{Experiment} \label{sec_2}

\begin{figure}[tb] 
    \centering
    \includegraphics[width=\columnwidth]{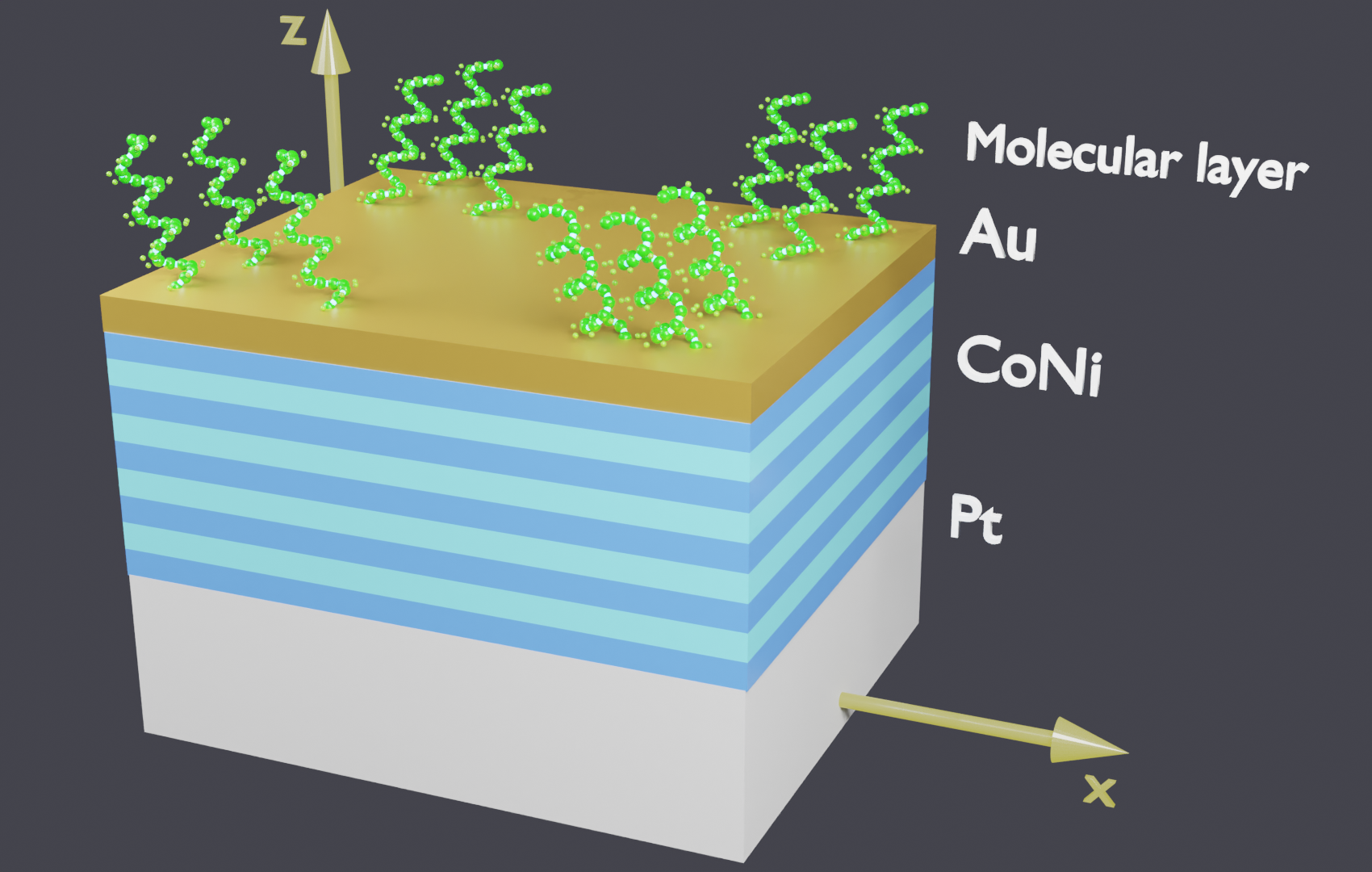}  
    \caption{Schematic representation of the sample layout. The molecular layer (greenish structures) is anchored on the CoNi-stack (light and dark blue) via an Au film. The CoNi multilayers are grown on a Pt buffer (grey), which is deposited on sapphire substrate with tantalum adhesion layer (not shown). The molecules are adsorbed under an angle with respect to the surface normal. The in-plane projections of the molecular axes are expected to be oriented randomly, so that their net component along the in-plane $x$-direction averages out. The axes projections on the direction perpendicular to the sample stack ($z$-direction) are, however, finite.}
    \label{fig:CoNi_slab} 
\end{figure}

\subsection{Sample preparation}

The Co/Ni multilayers (illustrated in Fig.\ \ref{fig:CoNi_slab}) with pronounced PMA were fabricated using direct current (dc) magnetron sputter deposition on c-plane sapphire, Al$_2$O$_3$(0001), with a thickness of 0.5~mm. The Ar pressure during sputtering was $3\times10^{-3}$~mbar and the base pressure of the system was $5\times10^{-9}$~mbar. 1.5~nm of Ta and 3~nm of Pt were used as adhesion layers and the sample was capped with 3~nm Au to prevent the Co/Ni multilayers from oxidation and enable the covalent attachment of the L- and D-oligopeptides. The full reference stack reads: Al$_2$O$_3$(0001)/Ta(15\,\AA)/Pt(30\,\AA)/[Co(2.5\,\AA)/Ni(6.5\,\AA)]$_3$/ Co(2.5\,\AA)/Au(30\,\AA) and will be referenced as FM throughout the whole paper. For further details on the sputtering processes, see \cite{Anna2026}.


5-(4-carboxyphenyl)-10,15,20-(triphenyl)porphyrin (TPCOOH) was purchased from PorphyChem. 
The synthesis of the chiral oligopeptide composed of alternating alanine (Ala) and 2-aminoisobutyric acid (Aib) residues with C-terminal cysteine (Ala-Aib)$_8$-Cys-OH as well as the N-terminal porphyrin chromophore was carried out by Peptide Specialty Laboratories GmbH. Such a sequence of the aminoacids assured $\alpha$-helical conformation of the oligopeptides in solutions used for fabrication of hybrid samples. 
Ethanol (HPLC grade) was purchased from Sigma-Aldrich and used with no further purifications. 

\begin{figure*}[t] 
    \centering
    \includegraphics[width=0.9\textwidth]{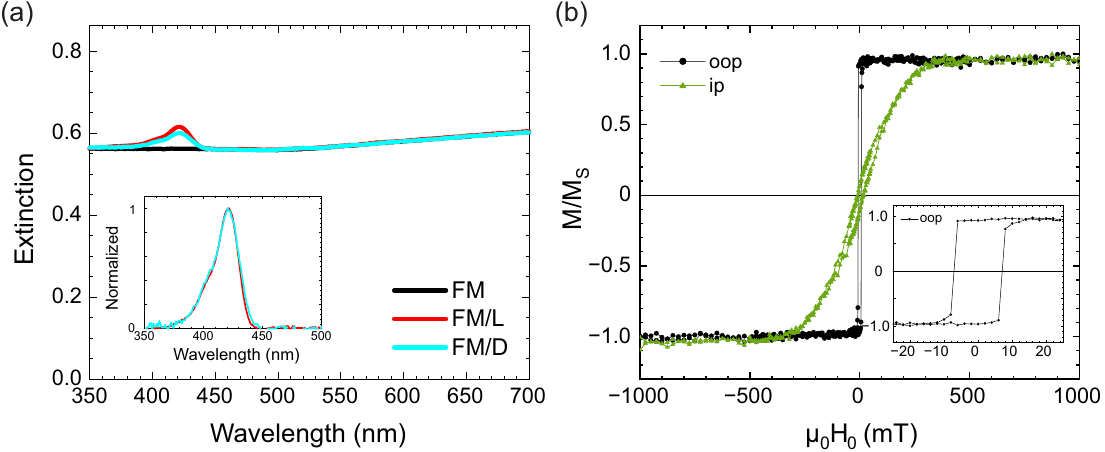}
    \caption{(a) Extinction spectra of the three samples FM, FM/L, and FM/D. The inset shows the corresponding normalized data. (b) Magnetization reversal loops for field out-of-plane (black) and in-plane (green) of an FM reference sample measured by VSM. }
    \label{fig:VSMFMR} 
\end{figure*}

The entire self-assembly of the chiral molecules on the sample surface was carried out under an Ar atmosphere. 
The FM stacks were immersed in 400~$\mu$L of $\approx90~\mu$M ethanolic solutions of L- and D-oligopeptides for 24 hrs.\ and subsequently thoroughly rinsed with ethanol and blow dried with N$_2$. Throughout this article we denote them as FM/L and FM/D, respectively.
To ensure identical concentrations of both enantiomers, the absorbances of their solutions at the Soret band (415~nm) were matched prior to immersion of the FM samples. 
The as-fabricated organic/inorganic hybrids were stored in the dark under vacuum. 

Note, that to assure highest accuracy during FMR measurements, first a bare FM sample, i.e.\ without molecules, denoted when necessary as FM$_\mathrm{L-ref}$ (FM$_\mathrm{D-ref}$) was measured. Subsequently, the same sample was used for adsorption of L-oligopeptides (D-oligopeptides) and re-measured as FM/L (FM/D), respectively.
The blank reference sample (without adsorbed molecules) was prepared by immersing it in neat ethanol for 24 hours.

As depicted in Fig.~\ref{fig:CoNi_slab}, the chiral molecules bind to the Au surface under a certain angle with respect to the surface normal. 
There is no preferential tilt-direction of their main axis. 
We thus assume that the spin-vector components of $\hat{\bm{\sigma}}$ \emph{within} the film plane average out, while there is a remaining component along the film normal ($z$-direction) only. 
For the detailed discussion on the structure of the molecular layer, please see \cite{Anna2026}. 

\subsection{Measurement techniques}

The extinction spectra of the samples were recorded in transmission mode using a Shimadzu SolidSpec-3700 UV-Vis-NIR spectrophotometer in the range of 350 to 700~nm.

We recorded the magnetization reversal loops using a commercial vibrating-sample magnetometer (VSM) from Microsense with in-plane (ip) and out-of-plane (oop) field direction.

We measured the FMR response of the samples on a broadband vector-network-analyzer (VNA) FMR setup in field-sweep mode at constant frequency. 
The samples were placed flip-chip on a coplanar waveguide and the absolute value of the microwave transmission parameter $\lvert S_{21}\rvert$ was used as the FMR-signal. 
Such field-sweeps were either measured for frequencies between 0.5~GHz and 35~GHz with the external magnetic field oriented either in the sample plane ($\theta_H=90$\textdegree) or along the film normal ($\theta_H=0$\textdegree). 
We fitted the FMR spectra with complex Lorentzians to retrieve the resonance field $\mu_0H_\mathrm{res}$ and peak-to-peak linewidth $\mu_0\Delta H_\mathrm{pp}$ for each frequency and angle, respectively. 
Finally, the magnetic anisotropy parameters were determined by fitting the resonance equation to the FMR data sets as derived in Sec.~\ref{sec_3}.

\subsection{Results and discussion}

\begin{figure*}[tb] 
    \centering
    \includegraphics[width=0.9\textwidth]{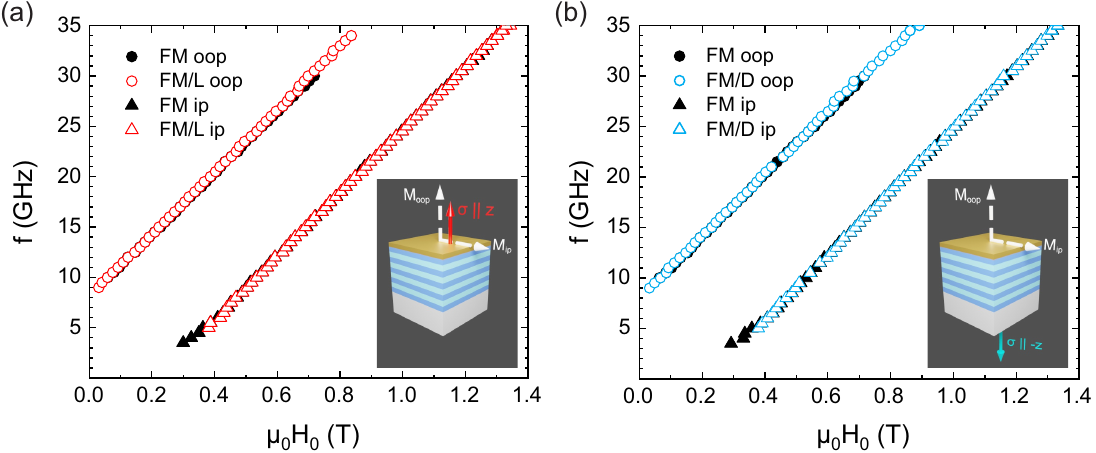}
    \caption{FMR frequency-field dependences for oop (circles) and ip (triangles) field direction of the samples with (a) L-enantiomer (red), (b) D-enantiomer (cyan), and the reference measurements without the enantiomers (black), respectively.}
    \label{fig:FMR} 
\end{figure*}

Figure \ref{fig:VSMFMR}(a) shows the extinction spectrum measured in transmission mode for a reference sample (black line) before adsorption of the molecules. The measured extinction (extinction = log$_{10}(1/T)$, where $T$ denotes transmittance), corresponds to the sum of absorption (intra- and interband transitions) and scattering losses of the thin metallic films, with absorption dominating over scattering. Elastic (Rayleigh) scattering is negligible because the studied films are continuous and exhibit a low surface roughness of typically $<$\,1\,nm. The transmission of light through the metallic heterostructures was measured again after chemisorption of the porphyrin-modified oligopeptides forming self-assembled monolayer (SAM) on the gold surface. 
The increase in extinction at the Soret band maximum ($\approx 420$~nm) upon SAM formation arises solely from light absorption by molecules covalently bound to the gold capping layer via side chain thiol group of C-terminal cysteine and serves as a fingerprint of the porphyrin macrocycle. 
This feature was therefore used to determine both the structure of the molecular layer and the surface concentrations of the enantiomers, yielding $5.3 \times 10^{13}$ molecules/cm$^{2}$ for the L- and $4.7 \times 10^{13}$ molecules/cm$^{2}$ for the D-enantiomer. 
For a detailed discussion concerning the origin of the observed extinction for a studied metallic stacks as well as the structure of the molecular layer and calculations of the molecular surface concentrations, the reader may refer to \cite{Anna2026}.

Figure \ref{fig:VSMFMR}(b) shows the magnetization reversal loops of a reference sample measured by VSM for two field directions, i.e.\ in-plane field (ip, i.e.\ parallel to the $x$-axis) and out-of-plane field (oop, i.e.\ parallel to the $z$-axis). 
The inset shows a magnification of the oop hysteretic part around zero field with a coercive field of $\mu_0H_\mathrm{c}=7$~mT.
The oop reversal loop shows a square hysteresis, which is a clear sign for the easy axis of the magnetic system being oriented along the film normal ($z$-axis) due to PMA. 

The FMR frequency-field dependence for the samples with/without the L- (red) and D-enantiomers (cyan) is given in Figs.~\ref{fig:FMR}(a) and \ref{fig:FMR}(b) for the ip- (triangles) and oop-field direction (circles), respectively. The insets show sketches of the relative orientation of the magnetization (white dashed arrows), determined by the direction of the external magnetic field, and the spin polarization $\hat{\bm \sigma}$ related to the molecular chirality, which was arbitrarily assigned to point away from the sample surface for the L-enantiomer and towards the sample surface for the D-enantiomer.

The frequency-field dependences of the FM/L and FM/D samples coincide with those of their respective reference samples. No significant deviations are observed, neither for the in-plane nor the out-of-plane magnetic field orientations.

From fitting these datasets with the resonance equations (\ref{eq:oop-resonance}) and (\ref{eq:ip-resonance}), that are derived in Sec.~\ref{sec_3}, we retrieve the effective magnetization $\mu_0M_\mathrm{eff}$ and magnetic anisotropy fields $\frac{2K_{2\perp}}{M_\mathrm{s}}$ and $\frac{2K_{4\perp}}{M_\mathrm{s}}$ as given in Table \ref{tab:fitparams}. Note, that as reported in literature \cite{CoNi_ML2020}, the higher order oop term $\propto \frac{2K_{4\perp}}{M_\mathrm{s}}$) has to be included, to appropriately fit the experimental data. 
The hybrid samples, FM/L and FM/D, could—within experimental error—be fitted using the same set of parameters as their respective reference samples without molecules, further confirming the absence of any effect of molecular chirality on the magnetization dynamics of the ferromagnet.

We note, that the comparatively large fourth-order perpendicular anisotropy observed in the CoNi multilayers cannot necessarily be interpreted as a simple higher-order correction to the same microscopic mechanism responsible for
the leading second-order perpendicular anisotropy. This is suggested by the fact that the two contributions favor different magnetic configurations: while the second-order term stabilizes the out-of-plane direction, the fourth-order contribution acts in favor of in-plane magnetization. 

\begin{figure*}[t] 
    \centering
    \includegraphics[width=0.9\textwidth]{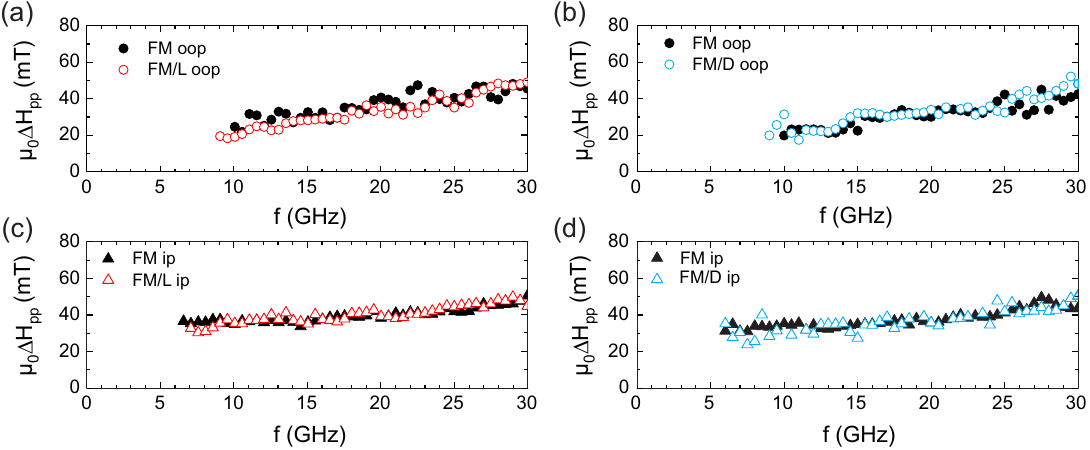}
    \caption {Frequency-dependence of the linewidth for field in (a,b) oop direction and (c,d) ip direction. Open symbols depict the samples with L-enantiomer (red) and D-enantiomer (cyan), whereas black depicts the reference sample.}
    \label{fig:FMRlinewidth} 
\end{figure*}

Besides film strain as origin for the fourth-order term, models based on
spatial fluctuations of the local perpendicular anisotropy, as originally discussed by Suhl \cite{suhl1957} and later extended by Slonczewski \cite{slonczewski1991}  for ultrathin magnetic films, may explain the observation. In such approaches, the local second-order anisotropy is assumed to vary spatially due to interface roughness, thickness fluctuations, strain variations, or electronic inhomogeneity. In the presence of exchange interaction, averaging over the resulting nonuniform magnetization texture
generates an effective higher-order anisotropy contribution. Importantly,
the resulting effective fourth-order term can acquire a sign opposite to
that of the leading perpendicular anisotropy contribution. Consequently, a
positive second-order perpendicular anisotropy may naturally coexist with a fourth-order term favoring in-plane magnetization.

Figure \ref{fig:FMRlinewidth} shows the frequency dependencies of the peak-to-peak resonance linewidths for the samples measured in both ip and oop magnetic field configurations, before (solid symbols) and after molecular adsorption (open symbols), respectively. 
No significant differences relative to the reference samples without molecules are observed. Consequently, the quantitative analysis was restricted to the reference samples. Linear fits to the data yield the Gilbert damping parameter, $\alpha$ from the slope and the inhomogeneous broadening $\mu_0\Delta H_{\mathrm{inh}}$ from the intercept, as summarized in Table \ref{tab:fitparams}. 

An observation that can be made upon careful inspection already for the reference samples, is a slight anisotropy of the Gilbert damping, with the value measured for magnetization oriented along the film normal $\alpha_\mathrm{oop}$, being larger than that for in-plane magnetization $\alpha_\mathrm{ip}$. This behavior becomes plausible when considering that a similar anisotropy is observed for the  $g$-factor, namely $g_\mathrm{oop}>g_\mathrm{ip}$. The anisotropy of the $g$-factor can be qualitatively understood within the framework of spin--orbit coupling (SOC). According to the model proposed by Bruno, in which SOC is treated as perturbation, the perpendicular magnetic anisotropy is directly related to the anisotropy of the orbital magnetic moment \cite{BrunoPhysRevB.39.865}:
\begin{align}
K_{2\perp} 
&\propto 
\Delta \mu_L = \mu_L^\mathrm{oop}-\mu_L^\mathrm{ip} .
\end{align}

Consequently, in PMA systems the orbital magnetic moment is expected to be slightly larger along the magnetic easy axis, i.e., $\mu_L^\mathrm{oop} > \mu_L^\mathrm{ip}$. Since the deviation of the $g$-factor from the free-electron value is proportional to the orbital moment \cite{KittelPhysRev1949, kittel_ferromagnetic_1951},
\begin{align}
g-2 
&\propto 
\mu_L ,
\end{align}
one correspondingly expects
\begin{align}
g_\mathrm{oop} > g_\mathrm{ip} .
\end{align}

Within the Kambersk\'y framework, the Gilbert damping is governed by SOC-mediated electronic transitions and approximately scales with the square of the SOC-induced orbital contribution \cite{mizukami_study_2001}:

\begin{align}
\alpha 
&\propto 
(g-2)^2 .
\end{align}

\noindent Therefore, the larger orbital contribution along the easy axis naturally leads to the expectation of a slightly enhanced damping for out-of-plane magnetization configurations, i.e., $\alpha_\mathrm{oop} > \alpha_\mathrm{ip}$.
\cite{BrunoPhysRevB.39.865, KittelPhysRev1949, kittel_ferromagnetic_1951, mizukami_study_2001}

\begin{figure*}[t] 
    \centering
    \includegraphics[width=0.8\textwidth]{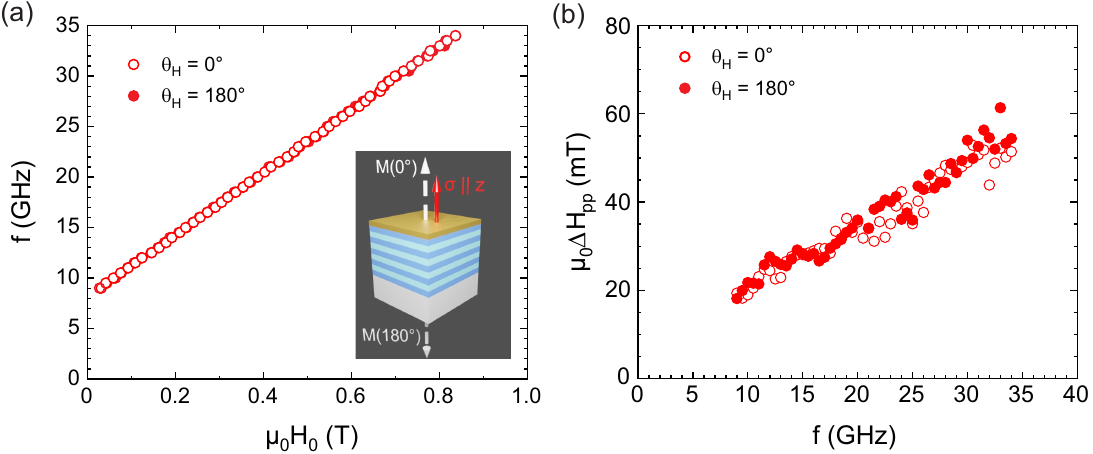}
    \caption{(a) FMR frequency-field dependence and (b) resonance linewidth for the the L-enatiomer and two field directions $\theta_H = 0$\textdegree (open symbols) and $\theta_H=180$\textdegree (filled symbols).}
    \label{fig:FMRfH} 
\end{figure*}

\begin{table*}[tb]
\caption{Fit parameters of the two reference samples: FM$_\mathrm{L-ref}$ and FM$_\mathrm{D-ref}$. The value for $\mu_0 M_\mathrm{s}$ is 0.9~T.}
\label{tab:fitparams}
\begin{ruledtabular}
\begin{tabular}{cccccccccc}
 sample & $\mu_0 M_{\mathrm{eff}}$ & $\frac{2K_{2\perp}}{M_\mathrm{s}}$   & $\frac{2K_{4\perp}}{M_\mathrm{s}}$  & $g_\mathrm{ip}$ & $g_\mathrm{oop}$ & $\alpha_\mathrm{ip}$ & $\alpha_\mathrm{oop}$ & $\mu_0\Delta H^\mathrm{ip}_{\mathrm{inh}}$ & $\mu_0\Delta H^\mathrm{oop}_{\mathrm{inh}}$  \\
 & (mT) & (mT) & (mT) & & & & & (mT) & (mT)\\
\hline
FM$_\mathrm{L-ref}$ & -334(8) & 1234(53) & -74(2) & 2.16(2) & 2.20(2) & 0.020(2) & 0.028(2) & 25(2) & 16(2)  \\
FM$_\mathrm{D-ref}$ & -327(9) & 1227(54) & -70(3) & 2.16(1) & 2.20(2) & 0.018(2) & 0.027(1) & 25(2) & 12(1) \\
\end{tabular}
\end{ruledtabular}
\end{table*}

The larger inhomogeneous linewidth broadening observed for in-plane magnetization can be qualitatively understood by considering a fixed distribution of local easy-axis orientations. Small local variations of the anisotropy-axis direction lead to local variations of the resonance field according to
\begin{align}
\Delta H_{\mathrm{inh}}
\propto
\left|
\frac{\partial H_\mathrm{res}}{\partial \theta_H}
\right|
\Delta \theta,
\end{align}
where $\Delta \theta$ describes the distribution of local easy-axis directions. The substantially stronger broadening observed for in-plane magnetization therefore indicates that the resonance field is considerably more sensitive to angular perturbations around the magnetic hard axis than around the easy axis.

In the data presented so far, the relative orientation between the magnetization and the spin polarization $\hat{\bm \sigma}$ attributed to the chirality of the molecular layer (parallel versus antiparallel alignment), was controlled by changing the molecular chirality, i.e., by comparing hybrid structures containing either the L- or D-enantiomer with their corresponding reference samples without molecules. However, for a hybrid structure with a fixed chirality, the relative orientation between $\hat{\bm \sigma}$ and the magnetization can also be reversed by changing the direction of the applied magnetic field from $\theta_H = 0$\textdegree \,($+z$-direction) to $\theta_H=180$\textdegree \,($-z$-direction), as schematically illustrated in Fig.~\ref{fig:FMRfH} together with corresponding frequency–field and linewidth–frequency dependencies for the hybrid structure with L-chirality. In line with the results presented in Fig.~\ref{fig:FMR} and Fig.~\ref{fig:FMRlinewidth}, reversing the direction of the applied magnetic field does not result in any shift of the resonance fields or change in the linewidth that could be correlated with the proximity of the chiral layer.

\section{Theoretical Description} \label{sec_3}

Due to the absence of an evidence of a significant effect in our FMR measurements, we decided to develop a theoretical description of the scenario. As we will see, this indeed provides insight into the nature of the CISS and MIPAC-effects in a manner consistent with our experimental findings.

For the description of magnetization dynamics--in particular FMR--it is useful to distinguish between equilibrium modifications of the magnetic free-energy landscape and non-equilibrium torque-driven effects. While both may lead to similar signatures in the equation of motion, their physical origin and dynamical consequences are fundamentally different.

MIPAC belongs to the equilibrium class. The adsorption of chiral molecules modifies the interfacial electronic structure and thereby the magnetic free-energy density $F(\bm M)$. This contribution enters via the effective field $\bm H_{\mathrm{eff}}$ and alters the energy landscape. Depending on the microscopic mechanism, it may appear, for example, as a Zeeman-like term
\begin{align}
F^{(Z)} = -\mu_0 \bm M \cdot \bm H_{\mathrm{chiral}},
\end{align}
or as an anisotropy contribution
\begin{align}
F^{(A)} = K_{\mathrm{chiral}} (\hat {\bm m} \cdot \hat {\bm n})^2,
\end{align}
with $\hat {\bm m}$ being the magnetization direction and $\hat {\bm n}$ set by the chiral interface. While the generation of a Zeeman-like field that introduces unidirectional properties can be related to local exchange interactions between the molecular layer and the thin film, the appearance of uniaxial anisotropy is based on local spin-orbit coupling between the orbitals of the molecules and the spins of the ferromagnet.

In contrast, CISS is intrinsically a non-equilibrium transport phenomenon. Charge carriers traversing or moving close to a chiral medium acquire a spin polarization $\hat{\bm \sigma}$ that is sustained by an external drive. This spin polarization enters the equation of motion as a torque, comprising both a field-like and a damping-like component. The latter enables energy transfer between the electronic and magnetic subsystems.

Although the formation of a chiral interface may involve transient non-equilibrium processes, no stationary spin polarization remains once equilibrium is reached. The effect is then fully captured by a modification of the free-energy landscape, without a continuous torque source.

The key distinction between MIPAC and CISS is therefore not the appearance of field-like terms, but whether the contribution derives from a free-energy functional or from an externally maintained spin polarization, as illustrated schematically in Fig.~\ref{fig:principle}. 

In FMR experiments, equilibrium contributions modify the resonance condition via changes in effective fields and anisotropies, leading to shifts of the resonance position. Non-equilibrium torque-driven effects, in contrast, act directly on the dynamics and can modify the linewidth and effective damping. Importantly, field-like torques may also shift the resonance position, so that frequency shifts alone do not uniquely identify equilibrium contributions.

In what follows, we assume ultrathin magnetic films and make use of a macrospin approach. Within this approximation, the magnetization is treated as a vector of fixed modulus, so that only its orientation remains dynamical. Longitudinal fluctuations are neglected, and the motion is therefore restricted to the surface of the unit sphere. It is then straightforward to introduce the normalized magnetization
\begin{equation}
\hat{\bm m} = \bm M / M_s, \qquad |\hat{\bm m}| = 1.
\end{equation}

\subsection{Equilibrium dynamics: MIPAC} \label{sec:4}

The impact of MIPAC--apart from the short time interval during which the molecules are adsorbed--manifests itself as reshaping the energy landscape, in which the magnetization vector moves. No external energy flow to or out of the system is present and thus all processes are determined by thermodynamics, i.e., can be described by the magnetic free energy. In the following, we consider the magnetic free-energy \emph{density} $F(\hat{\bm m})$, expressed as a function of the orientation of the unit vector $\hat{\bm m}$. A small variation $\delta \bm m$ changes the $F$ according to
\begin{equation}
\delta F = F(\hat{\bm m} + \delta \bm m) - F(\hat{\bm m}).
\end{equation}
For sufficiently small $\delta \bm m$, a Taylor expansion to first order yields
\begin{equation}
F(\hat{\bm m} + \delta \bm m)
=
F(\hat{\bm m})
+
\frac{\partial F}{\partial \hat{\bm m}} \cdot \delta \bm m
+
\mathcal{O}(|\delta \bm m|^2),
\end{equation}
so that
\begin{equation}
\delta F = \frac{\partial F}{\partial \hat{\bm m}} \cdot \delta \bm m,
\qquad
\frac{\partial F}{\partial \hat{\bm m}} \equiv \nabla_{\bm m} F.
\end{equation}
This motivates defining the effective magnetic field as
\begin{equation}
\bm H_{\mathrm{eff}} = -\frac{1}{\mu_0 M_s} \nabla_{\bm m} F,
\label{eq:effective_field}
\end{equation}
so that
\begin{equation}
\delta F = - \mu_0 M_s\, \bm H_{\mathrm{eff}} \cdot \delta \bm m.
\label{eq:deltaE_deltam}
\end{equation}

\noindent Thus, apart from the prefactor $-1/(\mu_0 M_s)$, 
$\bm H_{\mathrm{eff}}$ is given by the ordinary gradient 
of the free-energy density with respect to $\hat{\bm m}$ 
in $\mathbb{R}^3$. Here, only its projection onto the tangent plane of the unit sphere is dynamically relevant. Indeed, any allowed variation $\delta \bm m$ that preserves the normalization $\hat{\mathbf m}\cdot \hat{\mathbf m}=1$ must satisfy
\begin{equation}
\hat{\bm m} \cdot \delta \bm m = 0.
\end{equation}

In order to connect the effective field to magnetization dynamics, we use the vector identity 
\begin{equation}
\bm{a} \times (\bm{b} \times \bm{c})
= \bm{b}(\bm{a}\cdot\bm{c}) - \bm{c}(\bm{a}\cdot\bm{b}),
\label{eq:bac_cab}
\end{equation}
to obtain
\begin{equation}
-\hat{\bm m} \times (\hat{\bm m} \times \bm H_{\mathrm{eff}})
=
\bm H_{\mathrm{eff}} - (\hat{\bm m} \cdot \bm H_{\mathrm{eff}})\hat{\bm m}.
\end{equation}

\noindent The right-hand side corresponds to removing the component of $\bm H_{\mathrm{eff}}$ parallel to $\hat{\bm m}$. It is thus the projection of $\bm H_{\mathrm{eff}}$ onto the tangent plane of the unit sphere at $\hat{\bm m}$. 
Using $\bm H_{\mathrm{eff}} \propto -\partial F/\partial \hat{\bm m}$, this term describes a relaxation of the magnetization along the tangential component of the free-energy gradient, i.e.\ towards local minima of $F$ on the unit sphere, corresponding to dissipative relaxation towards equilibrium.

In contrast, the single cross product
\begin{equation}
\hat{\bm m} \times \bm H_{\mathrm{eff}}
\end{equation}
generates motion \emph{perpendicular} to the projected gradient of the free-energy density with respect to $\hat{\bm m}$ and corresponds to precessional dynamics. We finally note that from Eq.~(\ref{eq:bac_cab}) one obtains
\begin{align}
&\big[\hat{\bm m} \times (\hat{\bm m} \times \bm H_{\mathrm{eff}})\big]
\cdot
\big[\hat{\bm m} \times \bm H_{\mathrm{eff}}\big]
\nonumber\\
&=
(\hat{\bm m} \cdot \bm H_{\mathrm{eff}})\,
\hat{\bm m} \cdot (\hat{\bm m} \times \bm H_{\mathrm{eff}})
-
\bm H_{\mathrm{eff}} \cdot (\hat{\bm m} \times \bm H_{\mathrm{eff}})
= 0,
\end{align}
since $\hat{\bm m} \times \bm H_{\mathrm{eff}}$ is perpendicular to both $\hat{\bm m}$ and $\bm H_{\mathrm{eff}}$.

In consequence, away from stationary points, the vectors $\hat{\bm m} \times \bm H_{\mathrm{eff}}$ and $\hat{\bm m} \times (\hat{\bm m} \times \bm H_{\mathrm{eff}})$ span the tangent plane of the unit sphere. Any physically allowed dynamics of the magnetization can therefore be decomposed into a linear combination of these two contributions. This directly leads to the Landau--Lifshitz (LL) formulation, in which the magnetization dynamics is described by
\begin{equation}
\frac{d\hat{\bm{m}}}{dt}
= -\gamma\mu_0\, \hat{\bm{m}} \times \bm{H}_{\mathrm{eff}}
- \gamma\mu_0 \alpha\, \hat{\bm{m}} \times \big( \hat{\bm{m}} \times \bm{H}_{\mathrm{eff}} \big).
\label{eq:LL_equation}
\end{equation}

Effects on FMR, which occur in terms of the MIPAC mechanism can be absorbed into an effective field contribution. As mentioned earlier in the introduction, such a contribution may appear as an anisotropy field or a Zeeman-like directed field. These additional terms renormalize the free energy density, thereby creating, removing or redefining extrema, or influencing the slope and curvature of $F$. The MIPAC influence therefore does not actively drive the magnetization, but merely changes the energy surface, in which the magnetization dynamics takes place. This changes the FMR resonance condition and thereby the linewidth of the FMR signal, without, however, having impact on the damping rate of the magnetic system directly. The difference between damping rate and FMR linewidth will become evident in the following sections.

\subsection{Non-equilibrium dynamics: CISS} \label{sec_3_2}

The CISS effect generates a non-equilibrium spin polarization that can have an active influence on magnetization dynamics. It generates spin-torques that directly act on the magnetization vector. In our description, the spin-torque dynamics reduces to a flow on the unit sphere generated by tangent vector fields. We parameterize $\hat{\bm m}$ as
\begin{equation}
\hat{\bm{m}}(\theta,\varphi) =
\begin{pmatrix}
\sin\theta \cos\varphi \\
\sin\theta \sin\varphi \\
\cos\theta
\end{pmatrix},
\end{equation}
where $\theta$ is the polar angle measured with respect to the $z$-axis and $\varphi$ is the azimuthal angle measured from the $x$-axis (see Fig.~\ref{fig:CoNi_slab}). We further introduce a fixed spin-polarization vector $\hat{\bm{\sigma}}$, which is taken along the $-\hat{\bm{z}}$ direction,
\begin{equation}
\hat{\bm{\sigma}} = - \hat{\bm{z}}.
\end{equation}

\noindent The spin-torque vector field can, in general, be decomposed into two contributions: a damping-like (DL) and a field-like (FL) torque
\begin{equation}
\bm{\tau}(\hat{\bm{m}}) = c_\mathrm{DL}\,\bm{\tau}_\mathrm{DL} + c_\mathrm{FL}\,\bm{\tau}_\mathrm{FL},
\end{equation}
with
\begin{align}
\bm{\tau}_\mathrm{DL} &= -\,\hat{\bm{m}} \times (\hat{\bm{m}} \times \hat{\bm{\sigma}}), \label{eq:DL_torque} \\
\bm{\tau}_\mathrm{FL} &= \hat{\bm{m}} \times \hat{\bm{\sigma}}.
\label{eq:FL_torque}
\end{align}

\noindent Here, $\bm{\tau}_\mathrm{DL}$ and $\bm{\tau}_\mathrm{FL}$ are the dimensionless tangent vector fields on the unit sphere. The coefficients $c_\mathrm{DL}$ and $c_\mathrm{FL}$ therefore carry units of inverse time and set the strength of the corresponding torques in the equation of motion. The two torque fields are analyzed separately in the following.

\subsubsection{Damping-like torque}

Using the vector identity from Eq.~(\ref{eq:bac_cab}) on Eq.~(\ref{eq:DL_torque}), we get
\begin{equation}
\bm{\tau}_\mathrm{DL} = \hat{\bm{\sigma}} - \hat{\bm{m}}(\hat{\bm{m}}\cdot\hat{\bm{\sigma}}).
\end{equation}

\noindent For $\hat{\bm{\sigma}} = -\hat{\bm{z}}$, this becomes
\begin{equation}
\bm{\tau}_\mathrm{DL} = -\hat{\bm{z}} + \cos\theta\,\hat{\bm{m}}.
\end{equation}

\noindent In Cartesian components,
\begin{equation}
\bm{\tau}_\mathrm{DL} =
\begin{pmatrix}
\cos\theta \sin\theta \cos\varphi \\
\cos\theta \sin\theta \sin\varphi \\
-1 + \cos^2\theta
\end{pmatrix}.
\end{equation}

\noindent Its magnitude is $|\bm{\tau}_\mathrm{DL}| = \sin\theta$, so that one can write 
\begin{equation}
\bm{\tau}_\mathrm{DL} = \sin\theta \begin{pmatrix}
\cos\theta \cos\varphi \\
\cos\theta \sin\varphi \\
-\sin\theta
\end{pmatrix} 
= 
\sin\theta \, \mathbf{\hat{e}_{\theta}},
\label{eq:polar_tau_DL}
\end{equation}
where $\hat{\bm e}_{\theta}$ the polar unit vector. This implies that $\bm{\tau}_\mathrm{DL}$ is purely tangential,
\begin{equation}
\hat{\bm{m}} \cdot \bm{\tau}_\mathrm{DL} = 0.
\end{equation}

\noindent Geometrically, $\bm{\tau}_\mathrm{DL}$ is the tangential projection of $\hat{\bm{\sigma}}$ onto the sphere at the point $\hat{\bm m}$. It therefore drives motion along meridians toward the pole aligned with $\hat{\bm{\sigma}}$.

\subsubsection{Field-like torque}

For $\hat{\bm{\sigma}} = -\hat{\bm{z}}$ the field-like torque can be expressed in Cartesian components as
\begin{equation}
\bm{\tau}_\mathrm{FL} =
\begin{pmatrix}
-\sin\theta \sin\varphi \\
\sin\theta \cos\varphi \\
0
\end{pmatrix}.
\end{equation}

\noindent Its magnitude is $|\bm{\tau}_\mathrm{FL}| = \sin\theta$, which is equal to $|\bm{\tau}_\mathrm{DL}|$, and therefore 
\begin{equation}
\bm{\tau}_\mathrm{FL} = \sin\theta \begin{pmatrix}
-\sin\varphi \\
\cos\varphi \\
0
\end{pmatrix} 
= 
\sin\theta \, \mathbf{\hat{e}_{\varphi}},
\label{eq:polar_tau_FL}
\end{equation}

\noindent where $\hat{\bm e}_{\varphi}$ is the azimuthal unit vector. This implies that $\bm \tau_\mathrm{FL}$ is purely tangential,
\begin{equation}
\hat{\bm{m}} \cdot \bm{\tau}_\mathrm{FL} = 0.
\end{equation}

In contrast to $\bm{\tau}_\mathrm{DL}$, the field-like torque generates motion at constant $\theta$, corresponding to a rotation of $\hat{\bm{m}}$ around the $\hat{\bm{\sigma}}$ axis. Equivalently, it is tangent to circles of latitude on the sphere. 

\begin{figure}[tb] 
    \centering
    \includegraphics[width=\columnwidth]{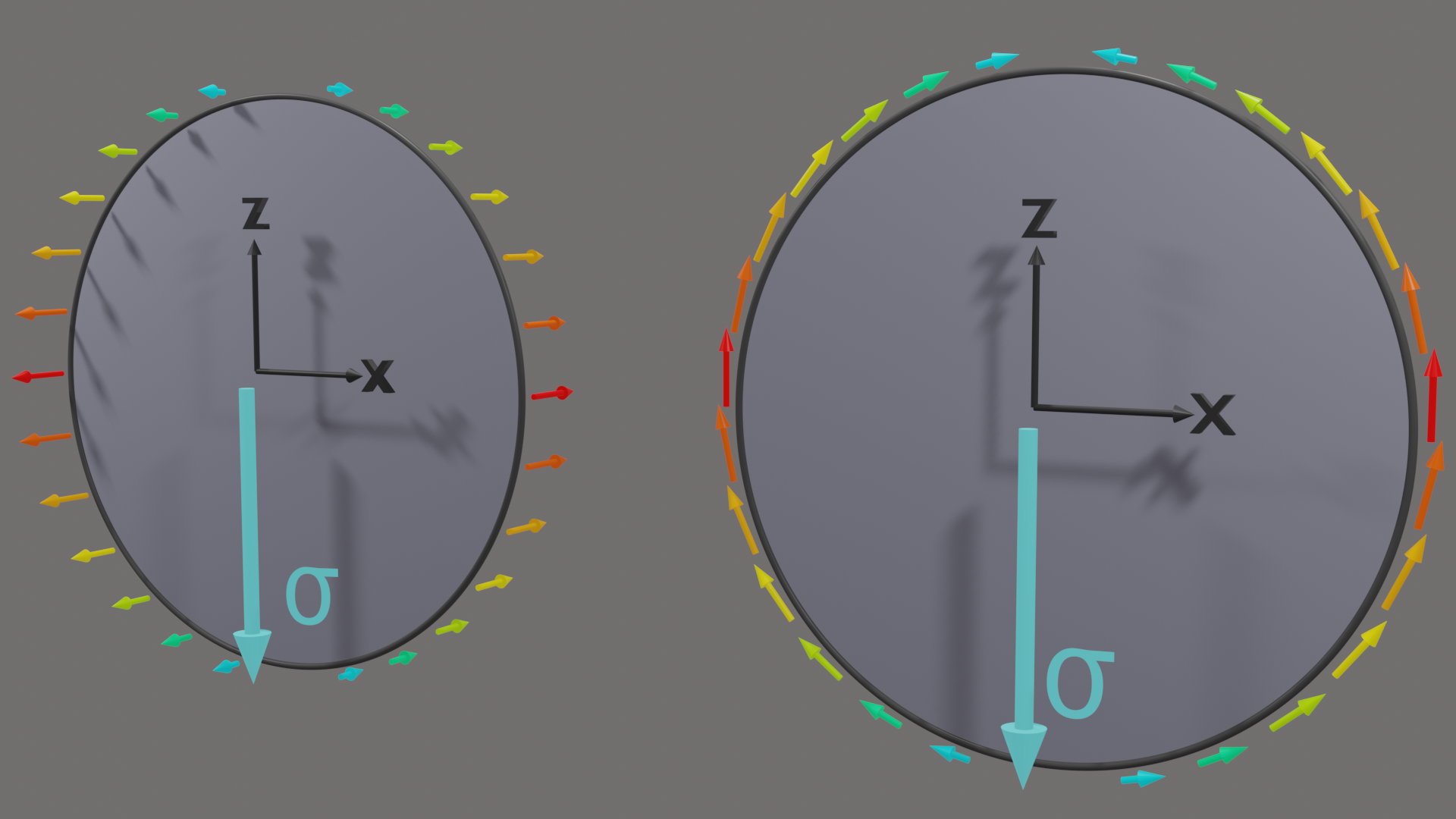}  
    \caption{Vector fields plotted on the $xz$ cross section of the unit sphere for the damping-like torque $\bm \tau_\mathrm{DL}$ (right circle) and field-like torque $\bm \tau_\mathrm{FL}$ (left circle). The magnitude of the torque vectors is color coded and also illustrated by the length of the arrows. The longest arrows, shown in red, correspond to the maximum magnitude $|\bm \tau_i| = 1$. The spin polarization $\hat{\bm \sigma}$ points towards the $-\hat{\bm z}$ direction.}
    \label{fig:vectorplot} 
\end{figure}

While the Cartesian components of $\bm{\tau}_\mathrm{DL}$ and $\bm{\tau}_\mathrm{FL}$ depend on $\varphi$, the polar forms of Eqs.~(\ref{eq:polar_tau_DL}) and (\ref{eq:polar_tau_FL}) show that both torque fields are rotationally symmetric about the $z$-axis. 
It is thus sufficient to visualize the two elementary torque fields on the unit \emph{circle}, as shown in Fig.~\ref{fig:vectorplot} . The torque magnitude is represented both by the arrow length and by the color coding, with red indicating maximum magnitude and cyan indicating minimum magnitude.

The geometric decomposition of the spin-torque directly determines how each torque modifies the LL formalism in Eq.~(\ref{eq:LL_equation}): the field-like torque is of the same form as the precessional term and can consequently be interpreted as arising from an additional effective field $\bm{H}_\mathrm{FL} \propto \hat{\bm{\sigma}}$. 

The damping-like torque has the same double-cross-product structure as the Landau--Lifshitz damping term, and enters the dissipative part.

\subsection{Influence of damping- and field-like torques on FMR eigenmodes}

Having established the geometric interpretation of the different torque contributions on the unit sphere, we now analyze their impact on the small-amplitude eigenmodes around a stationary magnetization direction that are probed by FMR. The resonance frequency and linewidth are determined by the eigenvalues of the linearized equation of motion. We therefore start from the Landau--Lifshitz equation including both damping-like and field-like spin-torque contributions,
\begin{align}
\frac{d\hat{\bm m}}{dt}
=
&-\gamma\mu_0\, \hat{\bm m} \times \bm H_{\mathrm{eff}}
- \gamma\mu_0 \alpha\, \hat{\bm m} \times (\hat{\bm m} \times \bm H_{\mathrm{eff}})
\nonumber\\
&+ c_\mathrm{DL}\,\bm\tau_\mathrm{DL}
+ c_\mathrm{FL}\,\bm\tau_\mathrm{FL},
\label{eq:LL_with_spin_torques}
\end{align}
with $\bm\tau_\mathrm{DL}$ and $\bm\tau_\mathrm{FL}$ defined in Eqs.~(\ref{eq:DL_torque}) and (\ref{eq:FL_torque}). Throughout this section, we assume a fixed spin polarization $\hat{\bm\sigma}$ perpendicular to the film plane.

\subsubsection{Stationary state and linearization}

The stationary magnetization direction $\hat{\bm m}_0$ is defined by the condition of vanishing total torque,
\begin{equation}
\left.\frac{d\hat{\bm m}}{dt}\right|_{\hat{\bm m}=\hat{\bm m}_0} = 0.
\label{eq:stationary_condition}
\end{equation}
In case of purely conservative dynamics, i.e.\ in the absence of spin torques, $\frac{d\hat{\bm m}}{dt}=0$ requires $\hat{\bm m} \times \bm H_{\mathrm{eff}}=0$, which enforces collinearity
between $\hat{\bm m}$ and $\bm H_{\mathrm{eff}}$.
Since $\bm H_{\mathrm{eff}} = -\frac{1}{\mu_0 M_s}\,\partial F/\partial \hat{\bm m}$, this condition is equivalent to $\partial F/\partial \hat{\bm m} \parallel \hat{\bm m}$. Geometrically, this implies that the tangential component of the gradient on the unit sphere vanishes, and the stationary state therefore corresponds to an extremum (minimum, maximum, or saddle point) of the free energy.

With spin-torque contributions, however, the situation must be distinguished more carefully. The field-like torque can be absorbed into an additional effective field and thus into a modified free-energy landscape. It may therefore change the stationary (equilibrium) direction, but only by shifting the position of the corresponding energy extremum.

The damping-like torque, in contrast, is nonconservative and cannot be written as $\bm A(\hat {\bm m}) = -\partial F/\partial \hat{\bm m}$. In its presence, the stationary state is generally determined by a genuine torque balance rather than by an extremum condition of the magnetic free energy alone. Accordingly, $\bm H_{\mathrm{eff}}$ need not be parallel to $\hat {\bm m}_0$. Since both damping-like and field-like contributions can enter the torque balance with finite tangential components, both can, in principle, shift the stationary orientation of $\hat{\bm m}$, albeit with a different physical meaning.

The occurrence of such a shift depends on the relative orientation of $\hat{\bm m}_0$ and $\hat{\bm\sigma}$. For the geometries considered below, where $\hat{\bm\sigma}$ is perpendicular to the film plane, two limiting cases are particularly relevant. For out-of-plane magnetization ($\hat{\bm m}_0 \parallel \hat{\bm\sigma}$) both damping-like and field-like torques vanish identically, so that the stationary direction is unaffected by the spin torques. By contrast, for in-plane magnetization ($\hat{\bm m}_0 \perp \hat{\bm\sigma}$) both torque terms are finite and can modify the stationary state.

In the following, we assume that $\hat{\bm m}_0$ denotes the actual stationary direction, including any possible shift induced by the spin torques. We then analyze the dynamics of small deviations
\begin{equation}
\hat{\bm m} = \hat{\bm m}_0 + \delta \bm m,
\qquad
|\delta \bm m| \ll 1,
\label{eq:m_expansion_main}
\end{equation}
with $\hat{\bm m}_0 \cdot \delta \bm m = 0$, i.e., the deviation is transverse to the stationary direction, as required by the normalization constraint. Linearizing the equation of motion yields
\begin{equation}
\frac{d}{dt}\,\delta \bm m
=
\hat{L}\,\delta \bm m,
\label{eq:linearized_eom_main}
\end{equation}
where the explicit evaluation of the linear operator $\hat{L}$ is given in Appendix~\ref{app:linearization_torques}. One finds that the linearized Landau--Lifshitz equation can be separated into equilibrium and non-equilibrium contributions,
\begin{align}
\frac{d}{dt}\,\delta \bm m
=
\left.\frac{d}{dt}\,\delta \bm m\right|_{\mathrm{eq}}
+
\left.\frac{d}{dt}\,\delta \bm m\right|_{\mathrm{neq}},
\end{align}
where the equilibrium part is given by
\begin{align}
\left.\frac{d}{dt}\,\delta \bm m\right|_{\mathrm{eq}}
=
&-\gamma\left(
\delta \bm m \times \bm H_{\mathrm{eff},0}
+
\hat{\bm m}_0 \times \delta \bm H_{\mathrm{eff}}
\right)
\nonumber\\
&+\gamma \alpha
\Big[
\delta \bm H_{\mathrm{eff}}
-
(\hat{\bm m}_0 \cdot \bm H_{\mathrm{eff},0})\delta \bm m
\nonumber\\
&-
(\delta \bm m \cdot \bm H_{\mathrm{eff},0})\hat{\bm m}_0
-
(\hat{\bm m}_0 \cdot \delta \bm H_{\mathrm{eff}})\hat{\bm m}_0
\Big],
\end{align}
and the non-equilibrium contribution reads
\begin{align}
\left.\frac{d}{dt}\,\delta \bm m\right|_{\mathrm{neq}}
=
&-c_\mathrm{DL}
\left[
(\hat{\bm m}_0 \cdot \hat{\bm\sigma})\delta \bm m
+
(\delta \bm m \cdot \hat{\bm\sigma})\hat{\bm m}_0
\right]
\nonumber\\
&+c_\mathrm{FL}\,\delta \bm m \times \hat{\bm\sigma}.
\end{align}

\noindent The equilibrium contribution is fully determined by the magnetic free-energy functional via $\bm H_{\mathrm{eff}} = -\frac{1}{\mu_0 M_s}\,\partial F/\partial \bm m$ and describes both precessional motion and intrinsic damping within the energy landscape, which may be influenced by variations of the effective field due to the MIPAC effect. In contrast, the non-equilibrium CISS contribution only arises from an externally maintained spin polarization $\hat{\bm \sigma}$ and enters the equation of motion as an additional torque term.

We note that while the field-like contribution formally resembles an effective field, it is not derived from the free energy of the isolated magnetic system and therefore remains a non-equilibrium term. The presence of $\hat{\bm \sigma}$ thus provides a direct criterion to identify non-equilibrium contributions at the level of the linearized dynamics.

\subsubsection{Local transverse dynamics}

As discussed in detail in Appendices~\ref{app:local_stiffness_fields} and \ref{app:projection_local_basis}, by introducing the local orthonormal basis $\{\hat{\bm e}_1,\hat{\bm e}_2,\hat{\bm e}_3\}$ with
$\hat{\bm e}_3=\hat{\bm m}_0$ and
\begin{equation}
\delta \bm m = m_1 \hat{\bm e}_1 + m_2 \hat{\bm e}_2,
\end{equation}
the linearized equation~(\ref{eq:linearized_eom_main}) reduces, after projection onto the transverse plane, to
\begin{equation}
\begin{pmatrix}
\dot m_1\\
\dot m_2
\end{pmatrix}
=
\gamma\mu_0
\begin{pmatrix}
-\alpha H_1 - \frac{c_\mathrm{DL}}{\gamma}\sigma_3
&
- H_2 + \frac{c_\mathrm{FL}}{\gamma}\sigma_3
\\
H_1 - \frac{c_\mathrm{FL}}{\gamma}\sigma_3
&
-\alpha H_2 - \frac{c_\mathrm{DL}}{\gamma}\sigma_3
\end{pmatrix}
\begin{pmatrix}
m_1\\
m_2
\end{pmatrix}.
\label{eq:dynamic_matrix_main}
\end{equation}
In this basis, the linear operator $\hat L$ restricted to the tangent plane is represented by the dynamic matrix. The conservative restoring contribution is determined by the curvature of the free-energy landscape around $\hat{\bm m}_0$. In the local transverse basis, this is captured by two so-called stiffness fields $H_1$ and $H_2$ (see Appendix~\ref{app:projection_local_basis}), defined as
\begin{equation}
H_i = \left.\frac{1}{\mu_0 M_s}\frac{d^2F}{dm_i^2}.
\right|_{\hat{\bm m}_0}.
\end{equation}
i.e., as the curvature of the free-energy density along the two transverse directions. Physically, the stiffness fields represent the restoring fields associated with small deflections and determine the ellipticity of the precession.

An important consequence is that only the longitudinal projection $\sigma_3=\hat{\bm m}_0\cdot\hat{\bm\sigma}$ enters the transverse linearized dynamics. The transverse components $\sigma_1$ and $\sigma_2$ do not appear in the projected $2\times2$ operator and therefore do not modify the FMR eigenvalues about a fixed stationary state; their role is instead to shift the stationary direction itself.

\subsubsection{Eigenvalues and physical interpretation}

Retaining only terms to first order in $\alpha$, $c_\mathrm{DL}$, and $c_\mathrm{FL}$, and neglecting products of these quantities, the eigenvalues of the dynamic matrix are
\begin{align}
\lambda
=
&-
\left[
\frac{\gamma\mu_0\alpha}{2}(H_1+H_2)
+
c_\mathrm{DL}\sigma_3
\right]
\nonumber\\
&\pm
i\left[
\gamma\mu_0\sqrt{H_1H_2}
-
\frac{c_\mathrm{FL}\sigma_3}{2}
\frac{H_1+H_2}
{\sqrt{H_1H_2}}
\right].
\label{eq:eigenvalues_main}
\end{align}
We thus identify the decay rate
\begin{equation}
\Gamma
=
\frac{\gamma\mu_0\alpha}{2}(H_1+H_2)
+
c_\mathrm{DL}\sigma_3,
\label{eq:gamma_main}
\end{equation}
and the resonance frequency
\begin{equation}
\omega
=
\gamma\mu_0\sqrt{H_1H_2}
-
\frac{c_\mathrm{FL}\sigma_3}{2}
\frac{H_1+H_2}
{\sqrt{H_1H_2}}.
\label{eq:omega_main}
\end{equation}
Here $\Gamma$ denotes the amplitude decay rate, such that the corresponding frequency-swept linewidth (full width at half maximum of the intensity) is $\Delta\omega_0=2\Gamma$.

These expressions clearly separate the roles of the two spin-torque components. The damping-like torque contributes exclusively to the symmetric (diagonal) part of the dynamic matrix and therefore modifies the decay rate, acting as damping or antidamping depending on the sign of $c_\mathrm{DL}\sigma_3$. By contrast, the field-like torque contributes to the antisymmetric (off-diagonal) part and therefore produces a purely reactive shift of the resonance frequency. Both effects are controlled solely by the projection $\sigma_3=\hat{\bm m}_0\cdot\hat{\bm\sigma}$, reflecting that only the component of the spin polarization \emph{parallel} to the stationary magnetization enters the transverse linearized dynamics directly. In Sec.~\ref{sec:8}, we evaluate the linewidth and resonance shift explicitly for the experimentally relevant thin-film geometries.

\subsection{Energy flow in the system}

The different roles of the torque contributions can be understood most clearly from their effect on the magnetic free-energy density $F(\hat{\bm m})$. According to Eq.~(\ref{eq:deltaE_deltam}), the time derivative of the energy follows from
\begin{equation}
\frac{dF}{dt}
=
\frac{\partial F}{\partial \hat{\bm m}}\cdot \frac{d\hat{\bm m}}{dt}
=
- \mu_0 M_s\, \bm H_\mathrm{eff}\cdot \frac{d\hat{\bm m}}{dt}.
\end{equation}

\subsubsection{Landau--Lifshitz field torque}

The precessional Landau--Lifshitz torque is given by
\begin{equation}
\left.\frac{d\hat{\bm m}}{dt}\right|_{\mathrm{prec}}
= -\gamma\, \hat{\bm m} \times \bm H_\mathrm{eff}.
\end{equation}

\noindent Its contribution to the energy change reads
\begin{align}
\left.\frac{dF}{dt}\right|_{\mathrm{prec}}
&=
- \mu_0 M_s\, \bm H_\mathrm{eff}\cdot
\left(-\gamma\, \hat{\bm m} \times \bm H_\mathrm{eff}\right) \nonumber \\
&= \gamma \mu_0 M_s\,
\bm H_\mathrm{eff}\cdot (\hat{\bm m} \times \bm H_\mathrm{eff}) = 0,
\end{align}
since $\bm H_\mathrm{eff}\cdot (\hat{\bm m} \times \bm H_\mathrm{eff}) = 0$. Thus, the Landau--Lifshitz field torque is strictly conservative: it generates motion along constant-energy trajectories on the unit sphere. This reflects the fact that $\bm H_\mathrm{eff}$ is derived from the free-energy functional.

\subsubsection{Field-like torque}

The field-like torque contributes
\begin{equation}
\left.\frac{d\hat{\bm{m}}}{dt}\right|_\mathrm{FL} = c_\mathrm{FL}\, \hat{\bm{m}} \times \hat{\bm{\sigma}}.
\end{equation}

\noindent This term is mathematically equivalent to precession in an additional effective field directed along $\hat{\bm\sigma}$. It can therefore be absorbed into the effective field and behaves like a conservative contribution. Equivalently, it may be associated with an additional Zeeman-like free-energy term. The field-like torque thus does not generate a net energy flow with respect to the corresponding total energy landscape, but modifies that landscape itself. In particular, it may shift the stationary direction by shifting the position of the free-energy extremum. This contribution thus can appear in similar manner as the MIPAC effect. This, however, vanishes when the external driving force establishing $\hat{\bm\sigma}$ is not maintained.   

\subsubsection{Landau--Lifshitz damping}

For the damping term, we obtain
\begin{align}
\left.\frac{dF}{dt}\right|_{\mathrm{LL}}
=
\gamma \alpha \mu_0 M_s\,
\bm H_\mathrm{eff}\cdot
\bigl[\hat{\bm m}\times(\hat{\bm m}\times \bm H_\mathrm{eff})\bigr].
\end{align}

\noindent Using Eq.~(\ref{eq:bac_cab}) we find
\begin{align}
\left.\frac{dF}{dt}\right|_{\mathrm{LL}}
=
\gamma \alpha \mu_0 M_s
\left[
(\hat{\bm m}\cdot \bm H_\mathrm{eff})^2
-
|\bm H_\mathrm{eff}|^2
\right].
\end{align}

\noindent Using \footnote{This identity follows directly from the geometric definitions $|\bm a \times \bm b| = |\bm a||\bm b|\sin\theta$ and 
$\bm a \cdot \bm b = |\bm a||\bm b|\cos\theta$, 
where $\theta$ is the angle between $\bm a$ and $\bm b$. 
Using $\sin^2\theta = 1 - \cos^2\theta$ immediately yields 
$|\bm a \times \bm b|^2 = |\bm a|^2 |\bm b|^2 - (\bm a \cdot \bm b)^2$.}
\begin{equation}
|\hat{\bm m}\times \bm H_\mathrm{eff}|^2
=
|\bm H_\mathrm{eff}|^2
-
(\hat{\bm m}\cdot \bm H_\mathrm{eff})^2,
\end{equation}
this becomes
\begin{equation}
\left.\frac{dF}{dt}\right|_{\mathrm{LL}}
=
-\gamma \alpha \mu_0 M_s\, |\hat{\bm m}\times \bm H_\mathrm{eff}|^2
\le 0.
\end{equation}

\noindent The Landau--Lifshitz damping term is strictly dissipative, always reducing the magnetic free-energy density.

\subsubsection{Damping-like torque}

For the damping-like contribution,
\begin{equation}
\left.\frac{d\hat{\bm m}}{dt}\right|_\mathrm{DL}
= c_\mathrm{DL}\big[-\hat{\bm m}\times(\hat{\bm m}\times\hat{\bm\sigma})\big],
\end{equation}
the energy change becomes
\begin{equation}
\left.\frac{dF}{dt}\right|_\mathrm{DL}
= -\mu_0 M_s c_\mathrm{DL}
\Big[
\bm H_{\mathrm{eff}}\cdot\hat{\bm\sigma}
-(\hat{\bm m}\cdot\hat{\bm\sigma})(\bm H_{\mathrm{eff}}\cdot\hat{\bm m})
\Big].
\end{equation}

\noindent This expression is nonzero in general. In particular, for
$\bm H_{\mathrm{eff}}\parallel\hat{\bm\sigma}$ one obtains
\begin{equation}
\left.\frac{dF}{dt}\right|_\mathrm{DL}
= -\mu_0 M_s c_\mathrm{DL}\sin^2\theta,
\end{equation}
which vanishes only at the poles. The damping-like torque is therefore intrinsically non-conservative and acts as a source or sink of magnetic free-energy density. In contrast to the field-like torque, it cannot be associated with motion along constant-energy trajectories.

In summary, the torque contributions fall into three distinct classes. The precessional Landau--Lifshitz term is genuinely conservative, whereas the field-like torque is conservative only at the level of the equations of motion: both can be absorbed into an effective field and thus correspond to dynamics on a (possibly modified) free-energy landscape. Nevertheless, the field-like torque ultimately originates from a non-equilibrium spin polarization maintained by external driving.

The Landau--Lifshitz damping term is dissipative and always reduces the magnetic free-energy density, driving the system towards local minima. In contrast to the field-like torque, the damping-like torque cannot be absorbed into $\bm H_\mathrm{eff}$. It induces a continuous energy flow during the precessional motion which provides either a source or a sink of energy. In the macrospin picture, this corresponds to a transition between precessional cones of different opening angles, i.e., a change of the orbit amplitude. As a consequence, the damping-like torque directly modifies the effective damping and can either enhance or compensate the intrinsic dissipation, while also affecting the dynamical response, including the resonance frequency.

\subsection{CISS in thin film FMR} \label{sec:8}

In the following, we discuss the relevant cases for our CoNi multilayer system. Figure~\ref{fig:sample_configurations} shows the configurations of normalized equilibrium magnetization $\hat{\bm m}_0$ and spin polarization $\hat{\bm \sigma}$ under consideration. As we have seen already, only the \emph{relative} orientation of these two quantities matters.

\subsubsection{Relation between LL damping and FMR linewidth}

At this point, we need to make a short statement about the connection between the intrinsic damping (rate) and the measured FMR linewidth. For a system governed by LL damping, according to Eqs.~(\ref{eq:gamma_main}) and (\ref{eq:omega_main}) the decay rate of the uniform mode in the linearized regime is
\begin{equation}
\Gamma = \frac{\gamma\mu_0\alpha}{2}\left(H_1+H_2\right),
\label{eq:Gamma_rate}
\end{equation}
while the resonance frequency reads
\begin{equation}
\omega = \gamma\mu_0\sqrt{H_1 H_2}.
\label{eq:res_condition_general}
\end{equation}
In general, $\Gamma$ is therefore not simply proportional to $\omega$, since it depends on the sum $H_1+H_2$, whereas $\omega$ depends on the geometric mean $\sqrt{H_1 H_2}$.

In typical FMR experiments the linewidth is measured in terms of the external magnetic field, i.e.\ $\Delta H_0$. The relation between frequency-swept linewidth (full width at half maximum of the intensity) $\Delta\omega = 2\Gamma$ and field-swept linewidths is
\begin{equation}
\Delta H_0 = \frac{\Delta\omega}{\left|\partial\omega/\partial H_0\right|}.
\label{eq:frequency_to_field_lw}
\end{equation}

For the uniform mode in the thin-film geometries considered below, the relevant geometric factors cancel in this conversion for the pure LL contribution. Using Eq.~(\ref{eq:omega_main}) without spin torques, the field derivative reads
\begin{align}
\frac{\partial\omega}{\partial H_0}
&= \frac{\gamma}{2\sqrt{H_1 H_2}}
\frac{\partial}{\partial H_0} \left(  H_1 H_2 \right) \nonumber \\
&= \frac{\gamma\mu_0}{2\sqrt{H_1 H_2}}
\left(H_1 + H_2\right),
\label{eq:omega_H0_partial}
\end{align}
such that the factor $(H_1 + H_2)$ cancels exactly. Here, the chain rule was used to evaluate 
$\partial (H_1 H_2)/\partial H_0$, assuming that the stiffness fields depend only explicitly on $H_0$, 
and not implicitly via the field dependence of the equilibrium magnetization $\bm m_0(H_0)$. This is equivalent to assuming that $\bm{m_0} \parallel \bm{H}_0$, i.e., dealing with an aligned FMR mode. One thus obtains
\begin{equation}
\mu_0\Delta H_0 = \frac{2\alpha}{\gamma}\,\omega,
\end{equation}
i.e.\ a linear dependence on frequency for the field-swept linewidth of the uniform mode in the absence of spin torques.


\subsubsection{Impact of CISS-related torque on different thin-film geometries}

\begin{figure}[tb] 
    \centering
    \includegraphics[width=0.4\columnwidth]
    {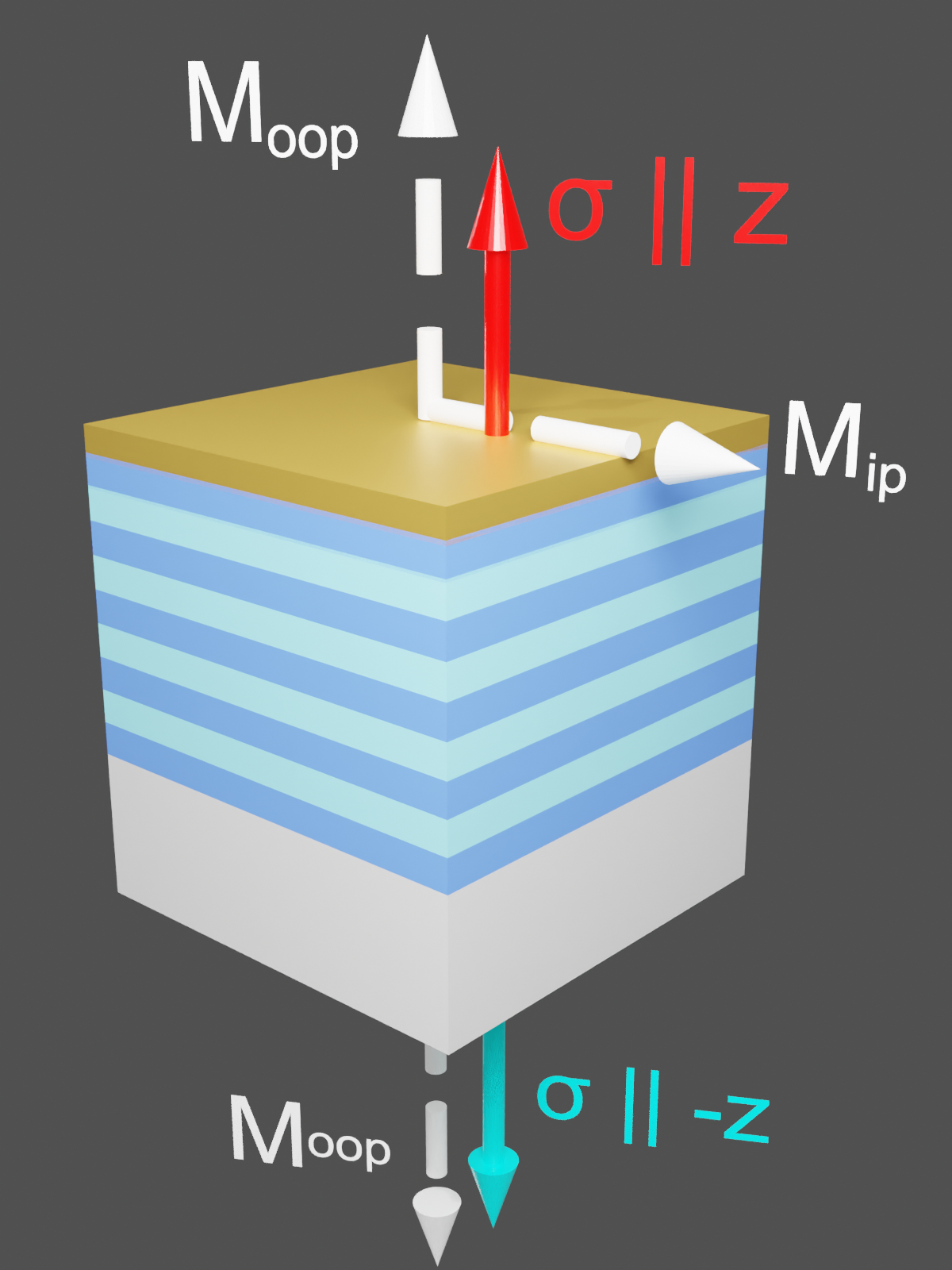}
    \caption{Schematic representation of different cases considered in the theoretical treatment. The spin polarization $\hat{\bm \sigma}$ is aligned perpendicular to the sample stack along either positive $z$-direction (red arrow) or negative $z$-direction (cyan arrow). The magnetization (white dashed arrows) is oriented perpendicular to the film stack--either parallel or antiparallel to the $z$--or within the film plane along the $x$-direction. Since the CoNi-multilayers exhibit negligible in-plane magnetic anisotropy, there is no preferential in-plane direction, which makes the choice of the $x$-axis arbitrary.}
    \label{fig:sample_configurations} 
\end{figure}

We now specialize Eq.~(\ref{eq:dynamic_matrix_main}) to thin-film geometries with either in-plane or out-of-plane equilibrium magnetization to analyze the impact of the field- and damping-like torques on the FMR response. The configurations considered are illustrated in Fig.~\ref{fig:sample_configurations}. 

For the CoNi multilayers considered here, the free-energy density is given by
\begin{align}
F &= -\mu_0 M_s H_0 (\bm{\hat{m}}\cdot\hat{\bm{h}}_0) 
+ \frac{\mu_0}{2} M_s M_{\mathrm{eff}} m_z^2 
- \frac{1}{2} K_{4 \perp} m_z^4 ,
\label{eq:free_energy_film}
\end{align}
with $\hat{\bm{h}}_0$ being the unit vector of the external field and 
\begin{equation}
\mu_0 M_{\mathrm{eff}} = \mu_0 M_s - \frac{2 K_{2 \perp}}{M_s}.
\end{equation}

Within this convention, a positive $K_{2\perp}$ favors an out-of-plane orientation of the magnetization, as it reduces the energy for finite $m_z$, while the demagnetizing term $\tfrac{\mu_0}{2}M_s^2 m_z^2$ penalizes it. The effective anisotropy is therefore governed by $M_{\mathrm{eff}}$, such that $M_{\mathrm{eff}} > 0$ corresponds to an in-plane easy axis and $M_{\mathrm{eff}} < 0$ to an out-of-plane easy axis. A positive $K_{4\perp}$ further stabilizes large $|m_z|$ within the present sign convention.

\medskip
\paragraph*{Out-of-plane magnetization} 

In this case $\hat{\bm m}_0 \parallel \hat{\bm z}$, while the transverse components are $m_x$ and $m_y$. Expanding the free energy to quadratic order yields identical stiffness fields (for details, see Appendix~\ref{app:stiffness_derivation}),
\begin{equation}
H_x = H_y
=
H_0 - M_{\mathrm{eff}} + \frac{2 K_{4 \perp}}{\mu_0 M_s}.
\end{equation}

The resonance frequency is therefore
\begin{equation}
\omega
=
\gamma\mu_0
\left(
H_0
-
M_{\mathrm{eff}}
+
\frac{2 K_{4 \perp}}{\mu_0 M_s}
\right).
\label{eq:oop-resonance}
\end{equation}

\medskip
\paragraph*{In-plane magnetization} 

Here $\hat{\bm m}_0 \parallel \hat{\bm x}$, and the transverse coordinates are $m_y$ and $m_z$, with equilibrium at $m_z=0$. The linearized dynamics is determined by the curvature of the free energy, i.e.\ by its second derivatives evaluated at equilibrium. For the quartic term $m_z^4$, the second derivative is proportional to $m_z^2$ and therefore vanishes at $m_z=0$. As a result, this term does not contribute to the transverse stiffness fields and does not enter the FMR frequency within the linear approximation. Evaluating the curvature yields
\begin{equation}
H_x = H_0,
\qquad
H_y = H_0 + M_{\mathrm{eff}}.
\label{eq:H_stiff_ip}
\end{equation}

The resonance frequency becomes
\begin{equation}
\omega
=
\gamma\mu_0
\sqrt{
H_0(H_0+M_{\mathrm{eff}})
}.
\label{eq:ip-resonance}
\end{equation}

In our case of CoNi multilayers with PMA, $M_{\mathrm{eff}}<0$, so that one can write 
\begin{align}
\sqrt{
H_0(H_0+M_{\mathrm{eff}})
}
=
H_0
\sqrt{
1-\frac{|M_{\mathrm{eff}}|}{H_0}
}.
\end{align}

A Taylor expansion for $H_0 \gg |M_{\mathrm{eff}}|$ yields
\begin{align}
\omega
\simeq
\gamma\mu_0
\left(
H_0
-
\frac{|M_{\mathrm{eff}}|}{2}
\right),
\end{align}
which shows that for a large external field, the frequency dependencies for in- as well as out-of-plane geometry exhibit a linear behavior with slope $\gamma\mu_0$. This is nicely seen in Fig.~\ref{fig:FMR}.

\subsubsection{Spin-polarization geometries}

The effect of the spin torque depends on the projection of $\hat{\bm \sigma}$ onto $\hat{\bm m}_0$:
\begin{itemize}
\item[(i)] $\hat{\bm \sigma} \parallel \hat{\bm m}_0$ (parallel),
\item[(ii)] $\hat{\bm \sigma} \parallel -\hat{\bm m}_0$ (antiparallel),
\item[(iii)] $\hat{\bm \sigma} \perp \hat{\bm m}_0$ (transverse).
\end{itemize}

These cases depicted in Fig.~\ref{fig:sample_configurations} are independent of whether the magnetization is oriented in- or out-of-plane.

\medskip
\paragraph*{Parallel and antiparallel configurations}

For $\hat{\bm \sigma} \parallel \pm \hat{\bm m}_0$, one has $\hat{\bm m}_0 \cdot \hat{\bm \sigma} = \pm 1$. In this case, both damping-like and field-like torques contribute to the linearized dynamics.

Using Eq.~(\ref{eq:gamma_main}) the decay rate becomes
\begin{equation}
\Gamma =
\frac{\gamma\mu_0\alpha}{2}
(H_x+H_y)
\pm c_\mathrm{DL},
\end{equation}
and the corresponding field-swept linewidth follows from Eqs.~(\ref{eq:frequency_to_field_lw}) and (\ref{eq:omega_H0_partial})
\begin{equation}
\mu_0\Delta H_0
=
\frac{
\gamma\mu_0\alpha(H_x+H_y)
\mp
2c_\mathrm{DL}
}
{
\left|\partial\omega/\partial H_0\right|
}
=
\frac{2 \alpha}{\gamma}\,\omega
\mp
\frac{2c_\mathrm{DL}}
{\left|\partial\omega/\partial H_0\right|}.
\label{eq:linewidth_DL_general}
\end{equation}

Equation~(\ref{eq:omega_main}) yields for the resonance frequency
\begin{equation}
\omega
=
\gamma\mu_0
\sqrt{H_xH_y}
\pm
\frac{c_\mathrm{FL}}{2}
\frac{
H_x+H_y
}{
\sqrt{H_xH_y}
}.
\end{equation}

Thus, the damping- and field-like torques produce a positive or negative linewidth or resonance-frequency offset, depending on the relative orientation between $\hat{\bm m}_0$ and $\hat{\bm \sigma}$.

\medskip
For the out-of-plane geometry with Eq.~(\ref{eq:oop-resonance}) one has
\begin{equation}
\frac{\partial\omega}{\partial H_0^{\mathrm{oop}}}
=
\gamma\mu_0,
\end{equation}
such that
\begin{equation}
\mu_0\Delta H^{\mathrm{oop}}_0
=
\frac{2 \alpha}{\gamma}\,\omega
\mp
\frac{2c_\mathrm{DL}}{\gamma}.
\label{eq:linewidth_oop_DL}
\end{equation}

Furthermore,
\begin{equation}
\omega^{\mathrm{oop}}
=
\gamma\mu_0
\left(
H_0
-
M_{\mathrm{eff}}
+
\frac{2 K_{4 \perp}}{\mu_0 M_s}
\right)
\pm
c_\mathrm{FL},
\label{eq:frequency_oop_FL}
\end{equation}
i.e., the damping- and field-like torques lead to constant linewidth and frequency offsets.

\begin{figure}[tb] 
    \centering
    \includegraphics[width=\columnwidth]
    {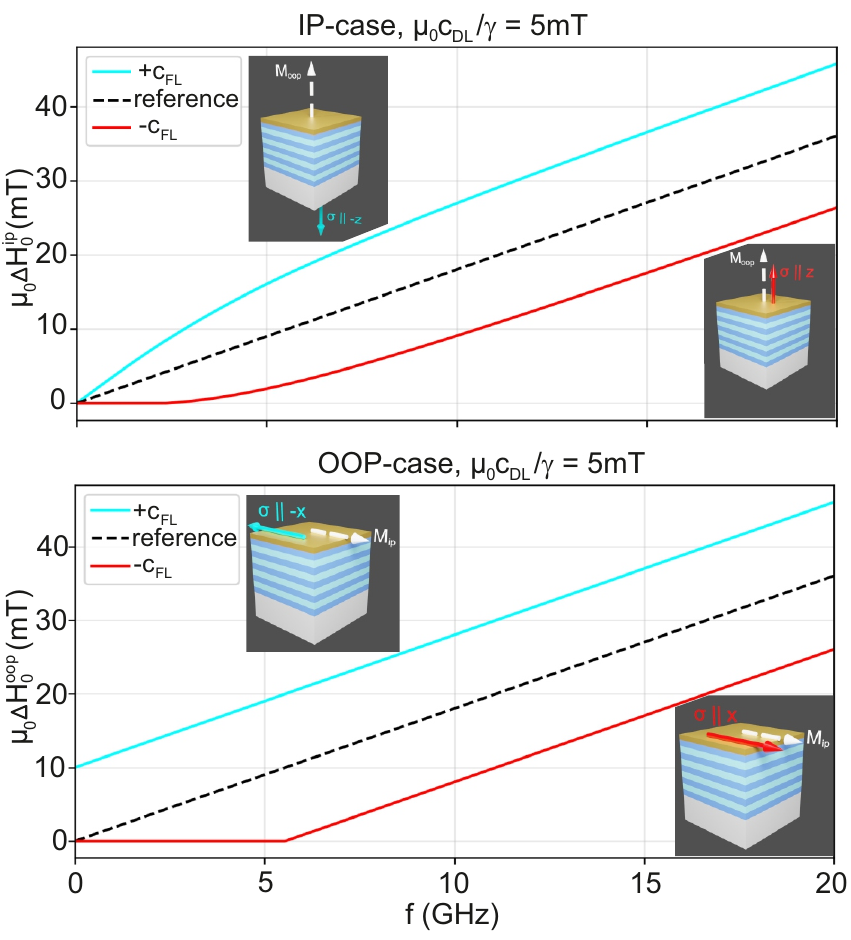}
    \caption{Calculated changes to the FMR linewidth of a thin film with CoNi parameters ($\mu_0 M_{\mathrm{eff}} = -330$~mT, $\frac{2K_{4\perp}}{M_\mathrm{s}} = -72$~mT) according to Eq.~(\ref{eq:linewidth_ip_DL}) for the in-plane configuration (upper panel) and according to Eq.~(\ref{eq:linewidth_oop_DL}) for the out-of-plane geometry (lower panel). The reference film for which no spin-polarization is assumed, i.e., $c_{\mathrm{DL}} = c_{\mathrm{FL}} = 0$ is shown as black dashed curve, the cyan and red curves show the dependence with spin-polarization $\hat{\bm \sigma}$ being either parallel or antiparallel to the static magnetization vector, respectively. The magnitude of the damping-like torque is $\mu_0 c_{\mathrm{DL}} / \gamma = 5$~mT. The inhomogenous linewidth offset $\Delta H_{\mathrm{inh}}$ has been omitted for clarity.}
    \label{fig:theory_plot_CISS_linewidth} 
\end{figure}

\medskip
For in-plane magnetization, both corrections become field dependent due to the ellipticity of the precession, but approach constant values in the high-field limit. For the in-plane geometry, using Eq.~(\ref{eq:ip-resonance}) one obtains
\begin{equation}
\frac{\partial \omega}{\partial H_0}
=
\frac{\gamma\mu_0(2H_0+M_{\mathrm{eff}})}
{2\sqrt{H_0(H_0+M_{\mathrm{eff}})}}.
\end{equation}
The field-swept linewidth, expressed in Tesla, therefore becomes
\begin{equation}
\mu_0\Delta H_0^{\mathrm{ip}}
=
\frac{2\alpha}{\gamma} \, \omega
\pm
\frac{
4c_{\mathrm{DL}}\sqrt{H_0(H_0+M_{\mathrm{eff}})}
}{
\gamma(2H_0+M_{\mathrm{eff}})
}.
\label{eq:linewidth_ip_DL}
\end{equation}

For large fields, $H_0 \gg |M_{\mathrm{eff}}|$, one finds
\begin{align}
\sqrt{H_0(H_0+M_{\mathrm{eff}})}
\simeq
H_0+\frac{M_{\mathrm{eff}}}{2},
\end{align}
and therefore
\begin{align}
\frac{
\sqrt{H_0(H_0+M_{\mathrm{eff}})}
}{
2H_0+M_{\mathrm{eff}}
}
\simeq
\frac{1}{2}.
\end{align}

\noindent Thus, Eq.~(\ref{eq:linewidth_ip_DL}) reduces to
\begin{equation}
\mu_0\Delta H^{\mathrm{ip}}_0
\simeq
\frac{2\alpha}{\gamma}\omega
\pm
\frac{2c_{\mathrm{DL}}}{\gamma},
\end{equation}
which is the same expression as for the out-of-plane case.

With Eqs.~(\ref{eq:H_stiff_ip}) and (\ref{eq:ip-resonance}) the corresponding resonance frequency reads
\begin{equation}
\omega^{\mathrm{ip}}
=
\gamma\mu_0\sqrt{H_0(H_0+M_{\mathrm{eff}})}
\pm
\frac{c_{\mathrm{FL}}}{2}
\frac{2H_0+M_{\mathrm{eff}}}
{\sqrt{H_0(H_0+M_{\mathrm{eff}})}}.
\label{eq:frequency_ip_FL}
\end{equation}
For the high-field limit, we now use
\begin{align}
\frac{
2H_0+M_{\mathrm{eff}}
}{
\sqrt{H_0(H_0+M_{\mathrm{eff}})}
}
\simeq
2,
\end{align}
such that
\begin{equation}
\omega_{\mathrm{ip}}
\simeq
\gamma\mu_0
\left(
H_0+\frac{M_{\mathrm{eff}}}{2}
\right)
\pm
c_{\mathrm{FL}}.
\end{equation}

\begin{figure}[tb] 
   \centering
  \includegraphics[width=\columnwidth]{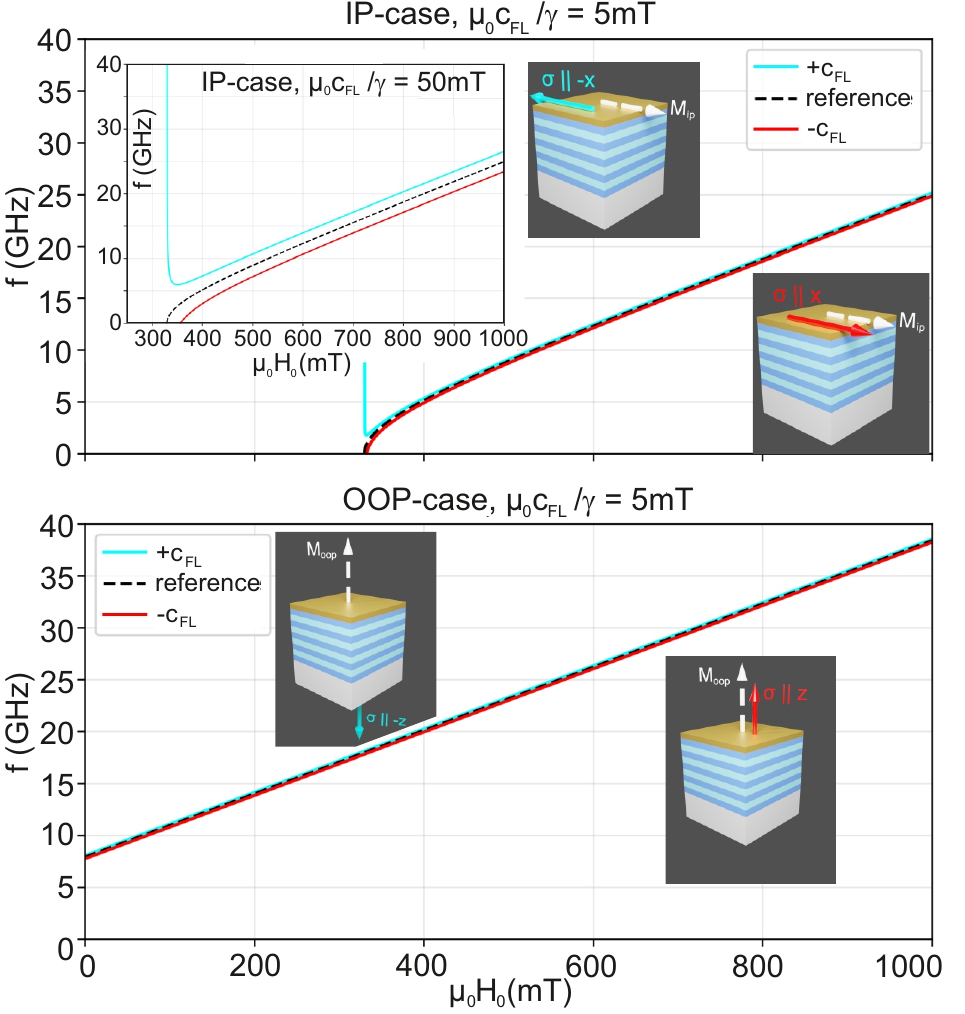}
\caption{Calculated changes to the FMR resonance condition of a thin film with CoNi parameters ($\mu_0 M_{\mathrm{eff}} = -330$~mT, $\frac{2K_{4\perp}}{M_\mathrm{s}} = -72$~mT) according to Eq.~(\ref{eq:frequency_ip_FL}) for the in-plane geometry (upper panel) and according to Eq.~(\ref{eq:frequency_oop_FL}) for the out-of-plane configuration (lower panel). The reference film for which no spin-polarization is assumed, i.e., $c_{\mathrm{DL}} = c_{\mathrm{FL}} = 0$ is shown as black dashed curve, the cyan and red curves show the dependence with spin-polarization $\hat{\bm \sigma}$ being either parallel or antiparallel to the static magnetization vector, respectively. The magnitude of the damping-like torque is $\mu_0 c_{\mathrm{DL}} / \gamma = 5$~mT. The inset in the upper panel shows the situation for the in-plane configuration, but with higher magnitude of the field-like torque, i.e., $\mu_0 c_{\mathrm{FL}} / \gamma = 50$~mT.}
\label{fig:theory_plot_CISS_resfield} 
\end{figure}

Figure~\ref{fig:theory_plot_CISS_linewidth} and Fig.~\ref{fig:theory_plot_CISS_resfield} show the response of the FMR linewidth and resonance condition to a CISS generated spin-polarization, respectively. The lower panels in Fig.~\ref{fig:theory_plot_CISS_linewidth} Fig.~\ref{fig:theory_plot_CISS_resfield} show the out-of-plane cases, while the upper panels show present the in-plane geometries. CoNi parameters from experiment as given in the figure captions have been used. The unperturbed situations without spin-polarization are shown by the black dashed curves. When a spin-polarization of magnitude $\mu_0 c_{\mathrm{DL/FL}} / \gamma = 5$~mT is added, the unperturbed response is up-shifted or down-shifted in frequency, depending on the sign of $c_{\mathrm{DL/FL}}$. This corresponds to adding a unidirectional character to the system, where linewidth and resonance condition are asymmetrically modified, depending on whether magnetization and spin-polarization are parallel or antiparallel, respectively. Note, that due to the larger field range of the resonance conditions shown in Fig.~\ref{fig:theory_plot_CISS_resfield} the shift due to the CISS-effect is less apparent. Therefore, an inset has been added for the in-plane case (upper panel of Fig.~\ref{fig:theory_plot_CISS_resfield}), where the magnitude of the spin-polarization was increased by a factor of ten, so that $\mu_0 c_{\mathrm{DL/FL}} / \gamma = 50$~mT.

The unidirectional character of the modification due to the CISS-torques can be used to distinguish it from a MIPAC related effect. In particular, the unidirectional influence of the FMR linewidth is a fingerprint of CISS. MIPAC influences mainly the resonance condition. Following the discussion of Sec.~\ref{sec:4}, when appearing as uniaxial anisotropy, MIPAC would lead to \emph{symmetric} changes of the resonance condition, i.e. have the same effect for either parallel or antiparallel orientation of the magnetization and spin-polarization. The only situation, where distinguishing CISS and MIPAC related effects needs a more careful inspection is, when the latter appears as directed Zeeman-like field. Similar to an exchange-bias contribution, this would generate unidirectional shifts of the resonance condition, too. However, upon removing the external driving force (e.g. electric current flow), CISS vanishes, while MIPAC would persist. That way, a separation remains possible.  

Finally, the upper panel of Fig.\ref{fig:theory_plot_CISS_resfield} shows a peculiar behavior of the resonance frequency, that occurs for the present CoNi system with PMA when measured along the in-plane hard direction. The resonance mode softens at the saturation field
\begin{align}
H_c = |M_{\mathrm{eff}}|.
\end{align}
The resonance frequency is then given by
\begin{align}
\omega_0
=
\gamma\mu_0
\sqrt{
H_0(H_0-H_c)
}.
\end{align}
Close to the saturation field,
\begin{align}
H_0-H_c
\rightarrow 0,
\end{align}
while the second factor $H_0$ remains finite. Consequently,
\begin{align}
\omega_0
\propto
\sqrt{H_0-H_c}.
\end{align}

\noindent The field-like correction term,
\begin{align}
\delta\omega_{\mathrm{FL}}
=
\pm
\frac{c_{\mathrm{FL}}}{2}
\frac{H_x+H_y}
{\sqrt{H_xH_y}},
\end{align}
therefore behaves as
\begin{align}
\delta\omega_{\mathrm{FL}}
\propto
\frac{c_{\mathrm{FL}}}
{\sqrt{H_0-H_c}},
\end{align}
and consequently diverges upon approaching the saturation field from above.

More generally, such a divergence occurs whenever the resonance frequency vanishes with square-root behavior near a critical field, i.e.\ when one stiffness field approaches zero while the second one remains finite.

\medskip
\paragraph*{Transverse configuration}

For $\hat{\bm \sigma} \perp \hat{\bm m}_0$, one has $\hat{\bm m}_0 \cdot \hat{\bm \sigma} = 0$. In this case, neither the field-like nor the damping-like torque contributes to the linearized eigenvalues and inewidth and resonance frequency reads like Eqs.~(\ref{eq:Gamma_rate}) and (\ref{eq:res_condition_general}), respectively. Thus, the linear FMR response remains unchanged. Physically, the torque shifts the equilibrium direction of the magnetization but does not modify the local curvature of the energy landscape around equilibrium. Consequently, the eigenfrequencies and damping rates remain unaffected to linear order.

\medskip
In summary, the linearized CISS-related FMR dynamics is governed by a clear separation of roles: the damping-like torque modifies the dissipative sector and thus the linewidth, whereas the field-like torque modifies the reactive sector and thus the resonance frequency. In both cases, only the longitudinal projection $\hat{\bm m}_0 \cdot \hat{\bm \sigma}$ enters the linearized dynamics, while transverse components affect only the equilibrium configuration.

\section{Conclusions}

A clear distinction between the CISS and MIPAC effects can be made in terms of equilibrium versus non-equilibrium contributions to magnetization dynamics. The essential distinction therefore lies in the physical origin rather than the mathematical form. Equilibrium contributions are fully captured by the free energy and act via effective fields, whereas non-equilibrium contributions arise from externally maintained spin polarizations and enter as torque terms in the equation of motion.

The MIPAC effect represents an equilibrium modification of the magnetic system. The adsorption of chiral molecules alters the interfacial electronic structure and thereby modifies the magnetic free-energy functional, $F(\hat{\bm m}) \rightarrow F(\hat{\bm m}) + \Delta F_{\mathrm{chiral}}(\hat{\bm m})$. This results in an additional effective field $\hat{\bm H}_{\mathrm{chiral}} = -\frac{\partial \Delta F_{\mathrm{chiral}}}{\partial \hat{\bm m}}$, which changes the energy landscape and equilibrium configuration without introducing a dynamical drive. Consequently, the resonance condition is altered, leading to shifts of the resonance field or resonance frequency. However, the contribution does not directly modify the dissipative dynamics and therefore does not intrinsically change the Gilbert damping or the FMR linewidth.

If the MIPAC contribution appears as a conventional uniaxial anisotropy,
\begin{align}
F^{A}
=
-K
(\hat{\bm m} \cdot \hat{\bm n})^2,
\end{align}
the resulting resonance-field shifts are symmetric under magnetization reversal $\hat{\bm m} \rightarrow -\hat{\bm m}$.
Such a signature can therefore be identified immediately as a static equilibrium anisotropy contribution due to MIPAC.

In contrast, a unidirectional equilibrium contribution,
\begin{align}
F^{Z}
=
-\mu_0 M_s H_{\mathrm{chiral}}
\,\hat{\bm m} \cdot \hat{\bm \sigma},
\end{align}
acts similarly to an exchange-bias-like effective field and produces asymmetric resonance-field shifts. Nevertheless, since this term still originates from the free energy, it remains conservative and therefore does not intrinsically modify the linewidth.

In contrast, the CISS effect is a non-equilibrium transport phenomenon. Charge carriers traversing a chiral medium acquire a spin polarization $\hat{\bm \sigma}$ sustained by an external drive, such as an electrical current. This spin polarization exerts spin torques on the magnetization $\hat{\bm m}$, including both field-like and damping-like contributions. The field-like component of this torque produces resonance-field shifts similar to an unidirectional effective field, whereas the damping-like component enables energy transfer between electronic and magnetic subsystems and can modify the effective damping, thereby changing the linewidth $\mu_0\Delta H_0$. This provides a particularly important experimental distinction: while both MIPAC and field-like CISS contributions may produce unidirectional resonance-field shifts, only a damping-like non-equilibrium torque generates a corresponding unidirectional linewidth modification. 

Importantly, a shift of the magnetic resonance condition alone does not provide unambiguous evidence for an equilibrium contribution, since non-equilibrium field-like torques can produce formally identical resonance shifts. A clear distinction can, however, be achieved by removing the external driving condition required to sustain the non-equilibrium CISS-related torques. If the resonance shift persists in the absence of the external drive, it must originate from an equilibrium MIPAC-type interaction.

Another point worth mentioning is, that a transient non-equilibrium regime may occur during molecular adsorption, where a temporary spin polarization can generate CISS-like torques. As the system relaxes, this contribution vanishes, leaving a modified equilibrium state described by an altered free energy, i.e., the MIPAC effect.

Finally, equilibrium modifications of the free-energy landscape may affect the measured FMR linewidth indirectly via changes in precession ellipticity. However, this does not correspond to a true change in damping. In the absence of a non-equilibrium spin source, no net energy transfer occurs over a precessional cycle, and the intrinsic damping remains unchanged. A genuine modification of damping requires a non-equilibrium torque, such as a damping-like contribution driven by $\hat{\bm \sigma}$.

Our comprehensive experimental study reveals no measurable influence of chiral molecules adsorbed on the surface of the magnetic heterostructure on its spin dynamics. A comparison between bare (reference) films and molecule-functionalized (hybrid) samples shows that both the resonance field and the linewidth remain unchanged. Furthermore, for hybrids containing either enantiomer, no expected asymmetry in the resonance signal was observed between the cases of spin polarization parallel and antiparallel to the magnetization, with the direction of spin polarization being dictated by the molecule chirality.

In light of our theoretical framework, which predicts chirality-induced modifications of magnetization dynamics via MIPAC-related contributions to the resonance field and linewidth, these results indicate that such effects are not detectable in the absence of sustained electron flow. We therefore conclude that, within the sensitivity of our measurements, there is no experimental evidence for a MIPAC-induced modification of spin dynamics in the studied ferromagnet. Importantly, this conclusion does not necessarily extend to the CISS effect, as all experiments reported here were conducted under conditions without sustained electron transport.

\subsection*{Author contributions} 
\noindent A.L.\ and J.L.\ conceived and supervised the work.
A.L., J.L.\ and K.L.\ wrote the paper with inputs from all authors. 
J.L., P.C.G., and R.G.\ developed and wrote the theory part.
A.L.,\ R.S.\ and O.H.\ prepared the samples. 
A.L., A.Se., A.Si. and K.L.\ performed and analyzed the FMR measurements, 
R.S.\ performed the VSM measurements, and A.L.A.\ measured the extinction spectra.
All authors read, commented, and revised the paper.

\subsection*{Funding}
\noindent This work is funded by the Deutsche Forschungsgemeinschaft (DFG, German Research Foundation) project no.\ 464974971. 
R.A.G.\ acknowledges support from FONDECYT grant 1250803 and Basal Program for Centers of Excellence, Grant CIA250002 (CEDENNA).
A.L.A.\ acknowledges financial support from the National Science Centre Poland (project no.\ 2020/39/I/ST5/00597).

\subsection*{Conflict of Interest}
\noindent The authors declare no conﬂict of interest.

\subsection*{Data Availability Statement}
\noindent The data of this study are available from the corresponding author upon reasonable request.

\appendix

\section{Linearization of torque terms in the Landau--Lifshitz equation}
\label{app:linearization_torques}

\noindent In this Appendix, we evaluate the linearized contributions of the individual torque terms entering the equation of motion. The linearization is performed with respect to the expansion
\begin{equation}
\hat{\bm m} = \hat{\bm m}_0 + \delta \bm m,
\qquad
|\delta \bm m| \ll 1,
\label{eq:m_expansion_appendix}
\end{equation}
where $\hat{\bm m}_0$ denotes the stationary magnetization direction. Since $|\hat{\bm m}|=1$, the deviation is restricted to the tangent plane,
\begin{equation}
\hat{\bm m}_0 \cdot \delta \bm m = 0.
\label{eq:transverse_constraint_appendix}
\end{equation}
For any torque contribution named $\bm T_i(\hat{\bm m})$ in the following, the linearized variation is defined as
\begin{equation}
\delta \bm T_i
=
\bm T_i(\hat{\bm m}_0 + \delta \bm m)
-
\bm T_i(\hat{\bm m}_0),
\label{eq:deltaT_definition_appendix}
\end{equation}
keeping only terms linear in $\delta \bm m$. Throughout, we use the bilinearity of the cross product and the expansion of the effective field
\begin{equation}
\bm H_{\mathrm{eff}}(\hat{\bm m}_0 + \delta \bm m)
=
\bm H_{\mathrm{eff},0}
+
\delta \bm H_{\mathrm{eff}},
\label{eq:Heff_expansion_appendix}
\end{equation}
with
\begin{equation}
\bm H_{\mathrm{eff},0}
=
\bm H_{\mathrm{eff}}(\hat{\bm m}_0).
\label{eq:Heff0_definition_appendix}
\end{equation}
The variation $\delta \bm H_{\mathrm{eff}}$ is itself linear in $\delta \bm m$ and will be specified below in the local tangent-plane basis. All effective fields and stiffness fields in this Appendix are magnetic fields in units of A/m. Hence the Landau--Lifshitz dynamics contains the factor $\gamma\mu_0$.

\subsubsection{Landau--Lifshitz precessional term}

For
\begin{equation}
\bm T_1 = -\gamma\mu_0\, \hat{\bm m} \times \bm H_{\mathrm{eff}},
\label{eq:T1_appendix}
\end{equation}
we obtain
\begin{align}
\bm T_1(\hat{\bm m}_0 + \delta \bm m)
&=
-\gamma\mu_0
(\hat{\bm m}_0 + \delta \bm m)
\times
(\bm H_{\mathrm{eff},0} + \delta \bm H_{\mathrm{eff}})
\nonumber\\
&=
-\gamma\mu_0\Big[
\hat{\bm m}_0 \times \bm H_{\mathrm{eff},0}
+
\delta \bm m \times \bm H_{\mathrm{eff},0}
\nonumber\\
&\hspace{1.5em}
+
\hat{\bm m}_0 \times \delta \bm H_{\mathrm{eff}}
+
\delta \bm m \times \delta \bm H_{\mathrm{eff}}
\Big].
\end{align}
Since the last term is of second order in $\delta \bm m$, it is neglected. Subtracting the equilibrium contribution
\begin{equation}
\bm T_1(\hat{\bm m}_0)
=
-\gamma\mu_0\,\hat{\bm m}_0 \times \bm H_{\mathrm{eff},0},
\end{equation}
yields
\begin{equation}
\delta \bm T_1
=
-\gamma\mu_0\left(
\delta \bm m \times \bm H_{\mathrm{eff},0}
+
\hat{\bm m}_0 \times \delta \bm H_{\mathrm{eff}}
\right).
\label{eq:LL_precession_term_linear}
\end{equation}
The first term describes the instantaneous rotation in the static effective field, while the second term contains the restoring contribution due to the variation of the effective field.

\subsubsection{Landau--Lifshitz damping}

For the damping term
\begin{equation}
\bm T_2
=
-\gamma\mu_0 \alpha\,
\hat{\bm m} \times (\hat{\bm m} \times \bm H_{\mathrm{eff}}),
\label{eq:T2_appendix}
\end{equation}
we use the vector identity
\begin{equation}
- \hat{\bm m} \times (\hat{\bm m} \times \bm H_{\mathrm{eff}})
=
\bm H_{\mathrm{eff}}
-
(\hat{\bm m} \cdot \bm H_{\mathrm{eff}})\hat{\bm m}.
\label{eq:bac_cab_appendix}
\end{equation}
Evaluating this expression at $\hat{\bm m}_0 + \delta \bm m$, we obtain
\begin{align}
\bm T_2(\hat{\bm m}_0 + \delta \bm m)
=
\gamma\mu_0 \alpha\Big[
\bm H_{\mathrm{eff},0}
+
\delta \bm H_{\mathrm{eff}}
\nonumber \\
-
\big(
(\hat{\bm m}_0 + \delta \bm m)
\cdot
(\bm H_{\mathrm{eff},0} + \delta \bm H_{\mathrm{eff}})
\big)
(\hat{\bm m}_0 + \delta \bm m)
\Big].
\end{align}
Expanding the scalar product gives
\begin{align}
(\hat{\bm m} \cdot \bm H_{\mathrm{eff}})
&=
(\hat{\bm m}_0 \cdot \bm H_{\mathrm{eff},0})
+
(\delta \bm m \cdot \bm H_{\mathrm{eff},0})
\nonumber\\
&\quad
+
(\hat{\bm m}_0 \cdot \delta \bm H_{\mathrm{eff}})
+
\mathcal{O}(\delta \bm m^2),
\end{align}
where the product $\delta \bm m \cdot \delta \bm H_{\mathrm{eff}}$ has been omitted since it is of second order. Keeping only linear terms, we find
\begin{align}
\bm T_2(\hat{\bm m}_0 + \delta \bm m)
&=
\gamma\mu_0 \alpha\Big[
\bm H_{\mathrm{eff},0}
+
\delta \bm H_{\mathrm{eff}}
\nonumber\\
&\quad
-
(\hat{\bm m}_0 \cdot \bm H_{\mathrm{eff},0})\hat{\bm m}_0
-
(\hat{\bm m}_0 \cdot \bm H_{\mathrm{eff},0})\delta \bm m
\nonumber\\
&\quad
-
(\delta \bm m \cdot \bm H_{\mathrm{eff},0})\hat{\bm m}_0
-
(\hat{\bm m}_0 \cdot \delta \bm H_{\mathrm{eff}})\hat{\bm m}_0
\Big].
\end{align}
Subtracting the equilibrium contribution
\begin{equation}
\bm T_2(\hat{\bm m}_0)
=
\gamma\mu_0 \alpha\Big[
\bm H_{\mathrm{eff},0}
-
(\hat{\bm m}_0 \cdot \bm H_{\mathrm{eff},0})\hat{\bm m}_0
\Big],
\end{equation}
we obtain
\begin{align}
\delta \bm T_2
&=
\gamma\mu_0 \alpha\Big[
\delta \bm H_{\mathrm{eff}}
-
(\hat{\bm m}_0 \cdot \bm H_{\mathrm{eff},0})\delta \bm m
\nonumber\\
&\quad
-
(\delta \bm m \cdot \bm H_{\mathrm{eff},0})\hat{\bm m}_0
-
(\hat{\bm m}_0 \cdot \delta \bm H_{\mathrm{eff}})\hat{\bm m}_0
\Big].
\label{eq:LL_damping_linear}
\end{align}

\subsubsection{Damping-like spin torque}

The damping-like torque reads
\begin{equation}
\bm\tau_\mathrm{DL}
=
\hat{\bm\sigma}
-
(\hat{\bm m} \cdot \hat{\bm\sigma})\hat{\bm m}.
\label{eq:tauDL_appendix}
\end{equation}
Since $\hat{\bm\sigma}$ is fixed, we have $\delta \hat{\bm\sigma}=0$. Evaluating at $\hat{\bm m}_0 + \delta \bm m$, we obtain
\begin{align}
\bm\tau_\mathrm{DL}(\hat{\bm m}_0 + \delta \bm m)
&=
\hat{\bm\sigma}
-
\Big[
(\hat{\bm m}_0 \cdot \hat{\bm\sigma})
+
(\delta \bm m \cdot \hat{\bm\sigma})
\Big]
(\hat{\bm m}_0 + \delta \bm m).
\end{align}
Keeping only terms linear in $\delta \bm m$, this becomes
\begin{align}
\bm\tau_\mathrm{DL}(\hat{\bm m}_0 + \delta \bm m)
&=
\hat{\bm\sigma}
-
(\hat{\bm m}_0 \cdot \hat{\bm\sigma})\hat{\bm m}_0
\nonumber\\
&\quad
-
(\hat{\bm m}_0 \cdot \hat{\bm\sigma})\delta \bm m
-
(\delta \bm m \cdot \hat{\bm\sigma})\hat{\bm m}_0.
\end{align}
Subtracting the equilibrium contribution
\begin{equation}
\bm\tau_\mathrm{DL}(\hat{\bm m}_0)
=
\hat{\bm\sigma}
-
(\hat{\bm m}_0 \cdot \hat{\bm\sigma})\hat{\bm m}_0,
\end{equation}
gives
\begin{equation}
\delta \bm\tau_\mathrm{DL}
=
-
\left[
(\hat{\bm m}_0 \cdot \hat{\bm\sigma})\delta \bm m
+
(\delta \bm m \cdot \hat{\bm\sigma})\hat{\bm m}_0
\right].
\label{eq:DL_linear_appendix}
\end{equation}

\subsubsection{Field-like spin torque}

For
\begin{equation}
\bm\tau_\mathrm{FL}
=
\hat{\bm m} \times \hat{\bm\sigma},
\label{eq:tauFL_appendix}
\end{equation}
we find
\begin{align}
\bm\tau_\mathrm{FL}(\hat{\bm m}_0 + \delta \bm m)
&=
(\hat{\bm m}_0 + \delta \bm m)\times \hat{\bm\sigma}
\nonumber\\
&=
\hat{\bm m}_0 \times \hat{\bm\sigma}
+
\delta \bm m \times \hat{\bm\sigma}.
\end{align}
Subtracting the equilibrium contribution yields
\begin{equation}
\delta \bm\tau_\mathrm{FL}
=
\delta \bm m \times \hat{\bm\sigma}.
\label{eq:FL_linear_appendix}
\end{equation}
This term has the same structure as the precessional torque, with the formal substitution $-\gamma\mu_0\,\bm H_{\mathrm{eff}}
\rightarrow c_\mathrm{FL}\,\hat{\bm\sigma}$.

Collecting Eqs.~(\ref{eq:LL_precession_term_linear}), (\ref{eq:LL_damping_linear}), (\ref{eq:DL_linear_appendix}), and (\ref{eq:FL_linear_appendix}), the linearized equation of motion takes the form
\begin{align}
\frac{d}{dt}\,\delta \bm m
=
&-\gamma\mu_0\left(
\delta \bm m \times \bm H_{\mathrm{eff},0}
+
\hat{\bm m}_0 \times \delta \bm H_{\mathrm{eff}}
\right)
\nonumber\\
&+\gamma\mu_0 \alpha
\Big[
\delta \bm H_{\mathrm{eff}}
-
(\hat{\bm m}_0 \cdot \bm H_{\mathrm{eff},0})\delta \bm m
\nonumber\\
&\qquad
-
(\delta \bm m \cdot \bm H_{\mathrm{eff},0})\hat{\bm m}_0
-
(\hat{\bm m}_0 \cdot \delta \bm H_{\mathrm{eff}})\hat{\bm m}_0
\Big]
\nonumber\\
&-c_\mathrm{DL}
\left[
(\hat{\bm m}_0 \cdot \hat{\bm\sigma})\delta \bm m
+
(\delta \bm m \cdot \hat{\bm\sigma})\hat{\bm m}_0
\right]
\nonumber\\
&+c_\mathrm{FL}\,\delta \bm m \times \hat{\bm\sigma}.
\label{eq:linearized_LL_full_appendix}
\end{align}

Equation~(\ref{eq:linearized_LL_full_appendix}) can be written in operator form as
\begin{equation}
    \frac{d}{dt}\delta \bm m
    =
    \hat L \, \delta \bm m ,
    \label{eq:linear_operator_appendix}
\end{equation}
where the linear operator $\hat L$ is defined by all terms in
Eq.~(\ref{eq:linearized_LL_full_appendix}) that are linear in $\delta\bm m$. Note, that $\delta \bm H_{\mathrm{eff}}$ is in fact linear in $\delta \bm m$, too, which directly follows from the effective field definition,
\begin{equation}
    \bm H_{\mathrm{eff}}
    =
    -\frac{1}{\mu_0 M_s}
    \frac{\partial F}{\partial \bm m}.
\end{equation}
Expanding the effective field around the equilibrium magnetization
$\hat{\bm m}_0$ according to $\bm m = \hat{\bm m}_0 + \delta \bm m$ one obtains to first order
\begin{equation}
    \bm H_{\mathrm{eff}}(\hat{\bm m}_0+\delta\bm m)
    \approx
    \bm H_{\mathrm{eff},0}
    +
    \left.
    \frac{\partial \bm H_{\mathrm{eff}}}{\partial \bm m}
    \right|_{\hat{\bm m}_0}
    \delta\bm m .
\end{equation}
Hence, the field variation becomes
\begin{equation}
    \delta \bm H_{\mathrm{eff}}
    =
    \bm H_{\mathrm{eff}}(\hat{\bm m}_0+\delta\bm m)
    -
    \bm H_{\mathrm{eff},0}
    \propto
    \delta\bm m .
\end{equation}

The vector expression of Eq.~(\ref{eq:linearized_LL_full_appendix}) is the explicit Cartesian representation of the abstract linearized operator used in the main text.

\section{Local basis and stiffness fields}
\label{app:local_stiffness_fields}

To describe transverse dynamics, we introduce a local orthonormal basis
\begin{equation}
\{\hat{\bm e}_1,\hat{\bm e}_2,\hat{\bm e}_3\},
\qquad
\hat{\bm e}_3 = \hat{\bm m}_0,
\end{equation}
where $\hat{\bm e}_1,\hat{\bm e}_2$ span the tangent plane perpendicular to the equilibrium magnetization. Since the magnetization is constrained by $|\bm m|=1$, its dynamics is restricted to the unit sphere. The local $\hat{\bm e}_3$ direction is therefore the surface normal of this sphere, while $\hat{\bm e}_1$ and $\hat{\bm e}_2$ span the local tangent plane. A small deviation is written as
\begin{equation}
\delta \bm m = m_1 \hat{\bm e}_1 + m_2 \hat{\bm e}_2.
\end{equation}

Because the magnetization moves on the unit sphere, the relevant energy curvature is the curvature along the constrained motion, i.e., the \emph{total} derivative
\begin{equation}
H_i = \frac{1}{\mu_0 M_s}\frac{d^2F}{dm_i^2}.
\end{equation}
The stiffness fields $H_i$ therefore have units of A/m.

We write the free energy as
\begin{equation}
F = F(m_1,m_2,m_3),
\qquad
m_3 = \sqrt{1-m_1^2-m_2^2}.
\end{equation}
Thus, $F$ depends on $m_i$ both explicitly and implicitly through $m_3(m_1,m_2)$. Applying the chain rule gives
\begin{equation}
\frac{dF}{dm_i}
=
\frac{\partial F}{\partial m_i}
+
\frac{\partial F}{\partial m_3}\frac{\partial m_3}{\partial m_i}.
\end{equation}
Differentiating once more yields
\begin{align}
\frac{d^2F}{dm_i^2}
&=
\frac{d}{dm_i}
\left(
\frac{\partial F}{\partial m_i}
+
\frac{\partial F}{\partial m_3}\frac{\partial m_3}{\partial m_i}
\right)
\nonumber\\
&=
\frac{\partial^2F}{\partial m_i^2}
+
2\frac{\partial^2F}{\partial m_i\partial m_3}\frac{\partial m_3}{\partial m_i}
+
\frac{\partial^2F}{\partial m_3^2}
\left(\frac{\partial m_3}{\partial m_i}\right)^2 \nonumber \\
& \quad +
\frac{\partial F}{\partial m_3}\frac{\partial^2 m_3}{\partial m_i^2}.
\label{eq:full_chain_rule_appendix}
\end{align}

\noindent The resulting expression can be interpreted as follows: the first term describes the curvature of $F$ with respect to $m_i$ at fixed $m_3$, the second and third terms arise from the mixed dependence of $F$ on $m_i$ and $m_3$, and the last term reflects the curvature of the constraint $m_3(m_1,m_2)$ associated with the unit-sphere condition. At equilibrium, $m_1=m_2=0$, and therefore
\begin{equation}
\left.\frac{\partial m_3}{\partial m_i}\right|_{\hat{\bm m}_0}=0.
\end{equation}
Hence, all terms proportional to $\partial m_3/\partial m_i$ vanish, and Eq.~(\ref{eq:full_chain_rule_appendix}) reduces to
\begin{equation}
\frac{d^2F}{dm_i^2}
=
\frac{\partial^2F}{\partial m_i^2}
+
\frac{\partial F}{\partial m_3}\frac{\partial^2 m_3}{\partial m_i^2}.
\label{eq:reduced_chain_rule_appendix}
\end{equation}

\noindent Using
\begin{equation}
m_3 = \sqrt{1-m_1^2-m_2^2}
\approx
1-\frac{1}{2}(m_1^2+m_2^2),
\end{equation}
one obtains
\begin{equation}
\left.\frac{\partial^2 m_3}{\partial m_i^2}\right|_{\hat{\bm m}_0}=-1.
\end{equation}

\noindent Moreover, since
\begin{equation}
\bm H_{\mathrm{eff}} = -\frac{1}{\mu_0 M_s}\frac{\partial F}{\partial \bm m},
\end{equation}
the derivative along the local normal direction is
\begin{equation}
\left.\frac{\partial F}{\partial m_3}\right|_{\hat{\bm m}_0}
= -\mu_0 M_s \bm H_{\mathrm{eff},0} \cdot \hat{\bm e}_3
= -\mu_0 M_s H_{\mathrm{eff},0},
\end{equation}
where $\bm H_{\mathrm{eff},0}=H_{\mathrm{eff},0}\hat{\bm e}_3$ is the equilibrium effective field.

\noindent Substituting these relations into Eq.~(\ref{eq:reduced_chain_rule_appendix}) gives
\begin{equation}
H_i
=
\frac{1}{\mu_0 M_s}\frac{d^2F}{dm_i^2}
=
\frac{1}{\mu_0 M_s}\frac{\partial^2F}{\partial m_i^2}
+
H_{\mathrm{eff},0}.
\label{eq:stiffness_decomposition_final}
\end{equation}

The stiffness fields naturally separate into two contributions. Because the effective field is defined as the gradient of the free-energy density, the first term describes the change of the effective field with magnetization. To first order, this variation is thus governed by the curvature of the free energy, i.e., its second derivative. One therefore obtains
\begin{equation}
\delta H_{\mathrm{eff},i}
=
-\,\kappa_i\, m_i,
\end{equation}
with
\begin{equation}
\kappa_i =
\frac{1}{\mu_0 M_s}
\left.
\frac{\partial^2 F}{\partial m_i^2}
\right|_{\hat{\bm m}_0}.
\end{equation}
The quantity $\kappa_i$ describes the Cartesian, or explicit, variation of the effective field with respect to a transverse magnetization component at fixed local normal direction. It should therefore not be identified with the full stiffness field. The full stiffness field $H_i=\kappa_i+H_{\mathrm{eff},0}$ also contains a geometric contribution arising from the rotation of the magnetization relative to the equilibrium field. At equilibrium,
\begin{equation}
\bm H_{\mathrm{eff},0} = H_{\mathrm{eff},0}\hat{\bm m}_0.
\end{equation}
A transverse deviation $\delta \bm m$ rotates the magnetization away from this direction. As a result, even a spatially uniform field $\bm H_{\mathrm{eff},0}$ acquires a transverse projection in the local frame, leading to a restoring contribution proportional to
\begin{equation}
H_{\mathrm{eff},0}\,\delta \bm m.
\end{equation}

\noindent Even if $\delta \bm H_{\mathrm{eff}} = 0$, a restoring force arises from the rotation of the magnetization relative to the equilibrium field. This originates from the geometric constraint that the motion is restricted to the curved surface of the sphere. As a result, a transverse displacement changes the local normal direction, so that the equilibrium field $\bm H_{\mathrm{eff},0}$ acquires a transverse component in the rotated frame.

Equation~(\ref{eq:stiffness_decomposition_final}) thus shows that the transverse stiffness fields consist of a Cartesian field variation and a geometric contribution associated with the constrained motion on the unit sphere.

\section{Projection of the torque terms onto the local tangent plane}
\label{app:projection_local_basis}

We evaluate the torque contributions in the local basis. The spin-polarization direction is decomposed as
\begin{equation}
\hat{\bm \sigma}
=
\sigma_1 \hat{\bm e}_1
+
\sigma_2 \hat{\bm e}_2
+
\sigma_3 \hat{\bm e}_3,
\qquad
\sigma_3 = \hat{\bm m}_0 \cdot \hat{\bm \sigma}.
\label{eq:sigma_local_basis_appendix}
\end{equation}

The linear variation of the effective field at fixed $m_3$, i.e., 
ignoring the constraint $|\bm m|=1$, introduced in Appendix~\ref{app:local_stiffness_fields}, is
\begin{equation}
\delta \bm H_{\mathrm{eff}}
=
- \kappa_1 m_1 \hat{\bm e}_1
- \kappa_2 m_2 \hat{\bm e}_2.
\label{eq:dHeff_local_appendix}
\end{equation}

\subsubsection{Landau--Lifshitz precessional term}

Starting from
\begin{equation}
\delta \bm T_1
=
-\gamma\mu_0
\left(
\delta \bm m \times \bm H_{\mathrm{eff},0}
+
\hat{\bm m}_0 \times \delta \bm H_{\mathrm{eff}}
\right),
\end{equation}
we evaluate the two contributions separately.

First, using $\bm H_{\mathrm{eff},0} = H_{\mathrm{eff},0}\hat{\bm e}_3$,
\begin{align}
\delta\bm m \times \bm H_{\mathrm{eff},0}
&=
H_{\mathrm{eff},0}
(m_1 \hat{\bm e}_1 + m_2 \hat{\bm e}_2)\times \hat{\bm e}_3
\nonumber\\
&=
H_{\mathrm{eff},0}(-m_1 \hat{\bm e}_2 + m_2 \hat{\bm e}_1).
\end{align}

\noindent Second,
\begin{align}
\hat{\bm e}_3 \times \delta\bm H_{\mathrm{eff}}
&=
- \kappa_1 m_1 \hat{\bm e}_2
+
\kappa_2 m_2 \hat{\bm e}_1.
\end{align}

\noindent Combining both terms yields
\begin{align}
\delta \bm T_1
&=
-\gamma\mu_0
\left[
(H_{\mathrm{eff},0}+\kappa_2) m_2 \hat{\bm e}_1
-
(H_{\mathrm{eff},0}+\kappa_1) m_1 \hat{\bm e}_2
\right].
\end{align}

\noindent Introducing the stiffness fields
\begin{equation}
H_i = \kappa_i + H_{\mathrm{eff},0},
\end{equation}
we obtain
\begin{align}
\delta \bm T_1
&=
-\gamma\mu_0
\left[
H_2 m_2 \hat{\bm e}_1
-
H_1 m_1 \hat{\bm e}_2
\right].
\end{align}

\noindent Hence,
\begin{equation}
\left(
\frac{d m_1}{dt}
\right)_{\mathrm{prec}}
=
-\gamma\mu_0 H_2 m_2,
\qquad
\left(
\frac{d m_2}{dt}
\right)_{\mathrm{prec}}
=
\gamma\mu_0 H_1 m_1.
\label{eq:prec_local_appendix}
\end{equation}

\subsubsection{Landau--Lifshitz damping term}

Starting from Eq.~(\ref{eq:LL_damping_linear}), we project onto the transverse plane. The longitudinal contributions vanish, yielding
\begin{equation}
\delta \bm T_2
=
\gamma\mu_0 \alpha
\left[
\delta \bm H_{\mathrm{eff}}
-
H_{\mathrm{eff},0}\,\delta \bm m
\right].
\end{equation}

\noindent Using Eq.~(\ref{eq:dHeff_local_appendix}) and $\delta \bm m = m_1 \hat{\bm e}_1 + m_2 \hat{\bm e}_2$, we obtain
\begin{align}
\delta \bm T_2
&=
-\gamma\mu_0 \alpha
\Big[
(\kappa_1 + H_{\mathrm{eff},0}) m_1 \hat{\bm e}_1
+
(\kappa_2 + H_{\mathrm{eff},0}) m_2 \hat{\bm e}_2
\Big].
\end{align}

\noindent Thus,
\begin{equation}
\left(
\frac{d m_1}{dt}
\right)_{\mathrm{LL}}
=
-\gamma\mu_0\alpha H_1 m_1,
\; \;
\left(
\frac{d m_2}{dt}
\right)_{\mathrm{LL}}
=
-\gamma\mu_0\alpha H_2 m_2.
\label{eq:LL_damping_local_appendix}
\end{equation}

\subsubsection{Damping-like torque in the local basis}

From Eq.~(\ref{eq:DL_linear_appendix}) we obtain
\begin{align}
\delta \bm\tau_\mathrm{DL}
&=
-
\left[
(\hat{\bm m}_0 \cdot \hat{\bm\sigma})\delta \bm m
+
(\delta \bm m \cdot \hat{\bm\sigma})\hat{\bm m}_0
\right].
\end{align}

\noindent After projection onto the tangent plane,
\begin{equation}
(\delta\bm\tau_\mathrm{DL})_\perp
=
-\sigma_3 \delta\bm m,
\end{equation}
which yields
\begin{equation}
\left(
\frac{d m_1}{dt}
\right)_\mathrm{DL}
=
- c_\mathrm{DL}\sigma_3\,m_1,
\quad
\left(
\frac{d m_2}{dt}
\right)_\mathrm{DL}
=
- c_\mathrm{DL}\sigma_3\,m_2.
\end{equation}

\subsubsection{Field-like torque in the local basis}

From
\begin{equation}
\delta \bm\tau_\mathrm{FL}
=
\delta \bm m \times \hat{\bm \sigma},
\end{equation}
we obtain
\begin{align}
\delta\bm m \times \hat{\bm \sigma}
&=
\sigma_3(-m_1 \hat{\bm e}_2 + m_2 \hat{\bm e}_1)
+
(m_1\sigma_2 - m_2\sigma_1)\hat{\bm e}_3.
\end{align}

\noindent After projection,
\begin{equation}
(\delta\bm\tau_\mathrm{FL})_\perp
=
\sigma_3(-m_1 \hat{\bm e}_2 + m_2 \hat{\bm e}_1).
\end{equation}

\noindent Thus,
\begin{equation}
\left(
\frac{d m_1}{dt}
\right)_\mathrm{FL}
=
c_\mathrm{FL}\sigma_3\,m_2,
\qquad
\left(
\frac{d m_2}{dt}
\right)_\mathrm{FL}
=
-\,c_\mathrm{FL}\sigma_3\,m_1.
\end{equation}

\section{Derivation of stiffness fields from the free-energy density}
\label{app:stiffness_derivation}

The stiffness fields are obtained from the curvature of the free-energy density on the unit sphere,
\begin{equation}
H_i = \frac{1}{\mu_0 M_s} \frac{d^2F}{dm_i^2}.
\end{equation}
Using the decomposition derived in Appendix~\ref{app:local_stiffness_fields}, this can be written as
\begin{equation}
H_i = \kappa_i + H_{\mathrm{eff},0},
\end{equation}
where $\kappa_i = \frac{1}{\mu_0 M_s} \frac{\partial^2 F}{\partial m_i^2}$ describes the intrinsic field response, while $H_{\mathrm{eff},0}$ accounts for the geometric contribution associated with the constraint $|\bm m|=1$. All stiffness fields are expressed in A/m; the corresponding field-like quantities in Tesla are obtained by multiplication with $\mu_0$.

In the following, we express the stiffness fields in the laboratory frame $(x,y,z)$, which coincides with the local basis at equilibrium.

We evaluate these expressions for the free-energy density
\begin{equation}
F =
-\mu_0 M_s H_0 (\bm{\hat{m}}\cdot\hat{\bm{h}}_0)
+\frac{\mu_0}{2} M_s M_{\mathrm{eff}}\, m_z^2
- \frac{1}{2} K_{4\perp} m_z^4,
\end{equation}
where
\begin{equation}
\mu_0 M_{\mathrm{eff}} = \mu_0 M_s - \frac{2K_{2\perp}}{M_s}.
\end{equation}

\subsubsection{Out-of-plane geometry}

For $\hat{\bm m}_0 \parallel \hat{\bm z}$, the transverse directions are $m_x$ and $m_y$.
The normalization constraint gives
\begin{equation}
m_z = \sqrt{1 - m_x^2 - m_y^2}
\;\approx\;
1 - \frac{1}{2}(m_x^2 + m_y^2),
\end{equation}
and
\begin{equation}
m_z^2 \approx 1 - (m_x^2 + m_y^2).
\end{equation}

Similarly, the fourth-order anisotropy term has to be
expanded to quadratic order as
\begin{equation}
m_z^4
=
\left(m_z^2\right)^2
\approx
\left[1-(m_x^2+m_y^2)\right]^2
\approx
1-2(m_x^2+m_y^2),
\end{equation}
where terms of fourth order in the transverse components
have been neglected. Substituting into $F$ and keeping only terms up to quadratic order (constant terms are omitted), we obtain
\begin{equation}
F^{(2)}
=
\frac{1}{2} \mu_0 M_s
\left[
H_0 - M_{\mathrm{eff}} + \frac{2K_{4\perp}}{\mu_0 M_s}
\right]
(m_x^2 + m_y^2).
\end{equation}

\noindent Thus,
\begin{equation}
H_x = H_y =
H_0 - M_{\mathrm{eff}} + \frac{2K_{4\perp}}{\mu_0 M_s},
\end{equation}
and the resonance frequency follows as
\begin{equation}
\omega
=
\gamma\mu_0 \sqrt{H_x H_y}
=
\gamma\mu_0
\left(
H_0 - M_{\mathrm{eff}} + \frac{2K_{4\perp}}{\mu_0 M_s}
\right).
\end{equation}

\subsubsection{In-plane geometry}

For $\hat{\bm m}_0 \parallel \hat{\bm x}$, the transverse directions are $m_y$ and $m_z$.
The normalization constraint gives
\begin{equation}
m_x = \sqrt{1 - m_y^2 - m_z^2}
\;\approx\;
1 - \frac{1}{2}(m_y^2 + m_z^2).
\end{equation}

Substituting into $F$ and keeping quadratic terms yields
\begin{equation}
F^{(2)}
=
\frac{1}{2} \mu_0 M_s H_0\, m_y^2
+
\frac{1}{2} \mu_0 M_s (H_0 + M_{\mathrm{eff}})\, m_z^2.
\end{equation}

\noindent Thus,
\begin{equation}
H_y = H_0,
\qquad
H_z = H_0 + M_{\mathrm{eff}},
\end{equation}
and the resonance frequency becomes
\begin{equation}
\omega
=
\gamma\mu_0 \sqrt{H_y H_z}
=
\gamma\mu_0 \sqrt{H_0 \left(H_0 + M_{\mathrm{eff}}\right)}.
\end{equation}

The fourth-order anisotropy contributes only in the out-of-plane geometry, where $m_z=1$ in equilibrium. For in-plane magnetization, it does not enter the quadratic expansion and therefore does not affect the linearized FMR response.


\bibliography{CISS_effect_final.bib}

\end{document}